\documentclass[pdftex,twocolumn,3]{jour3}          

\usepackage[utf8]{inputenc}
\usepackage{epsf}
\usepackage{latexsym,amssymb,euscript}
\usepackage{widetext}
\usepackage[dvips]{graphicx}
\usepackage[numbers,sort&compress]{natbib}
\usepackage{amsmath}
\usepackage{nicefrac}
\usepackage{slashed}
\usepackage{booktabs}
\usepackage{hyperref}
\usepackage{braket}
\usepackage{chngcntr}
\usepackage{bm}
\usepackage{bbold}
\usepackage{graphics}
\usepackage{graphicx}
\usepackage{mciteplus}
\usepackage{bbold}
\usepackage{pdfpages}
\usepackage[titletoc]{appendix}

\usepackage{ulem}
\usepackage{microtype} 

\graphicspath{{./figures/}}
\hypersetup{
 linktocpage = true,
 urlcolor = urlblue,
 colorlinks = true,
 linkcolor = urlblue,
 anchorcolor = urlblue,
 citecolor = urlblue,
 pdfstartview = {XYZ null null 1.25} 
           }
\usepackage[left=2cm, right=2cm]{geometry}
\usepackage{pstricks}
\usepackage{color}
\usepackage{xcolor}
\definecolor{urlblue}{rgb}{0.2,0.4,0.7}
\definecolor{citegreen}{rgb}{0,0.4,0.2}
\definecolor{linkred}{rgb}{0.9,0.2,0.1}
\usepackage{float}
\usepackage{academicons}
\definecolor{orcidlogocol}{HTML}{A6CE39}
\usepackage{fancyhdr}
\newcommand{\drv}{{\rm d}}
\newcommand{\prodscal}{\text{\large{\textbf{\textperiodcentered}}}}

\newcommand{\as}{\alpha_s}
\newcommand{\LQCD}{\Lambda_{\rm QCD}}
\newcommand{\MSb}{\overline{\rm MS}}

\newcommand{\LL}{{\rm LL/LO}}

\newcommand{\NLLp}{{\rm NLL/NLO^+}}
\newcommand{\CnLL}{{\cal C}_n^\LL}

\newcommand{\CnNLLp}{{\cal C}_n^\NLLp}

\newcommand{\DY}{\Delta Y}
\newcommand{\JPsi}{J/\psi}

\newcommand{\qTab}{\bm{q_{T_{1,2}}}}
\newcommand{\qTa}{\bm{q_{T_1}}}
\newcommand{\qTb}{\bm{q_{T_2}}}
\newcommand{\qT}{\bm{q_T}}

\newcommand{\tcite}[1]{~\cite{#1}}
\newcommand{\tref}[1]{~\ref{#1}}
\newcommand{\eref}[1]{~\eqref{#1}}

\newcommand{\tarr}{
\begin{array}}
\newcommand{\earr}{\end{array}}

\newcommand{\orcidFGC}{\href{https://orcid.org/0000-0003-3299-2203}{\includegraphics[scale=0.1]{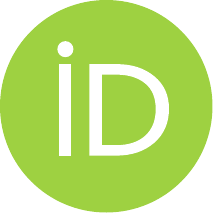}}}

\newcommand{\orcidAP}{\href{https://orcid.org/0000-0001-8984-3036}{\includegraphics[scale=0.1]{figures/logo-orcid.pdf}}}

\smartqed  

\journalname{}

\begin{document}

\normalem

\title{Mueller-Navelet jets at the LHC: Hunting data with azimuthal distributions
}

\subtitle{}

\author{
Francesco Giovanni Celiberto
\thanksref{e1,addr1,addr2,addr3} \orcidFGC
\and
Alessandro Papa
\thanksref{e2,addr4,addr5} \orcidAP
}

\thankstext{e1}{{\it e-mail}:
\href{mailto:fceliberto@ectstar.eu}{fceliberto@ectstar.eu} (corresponding author)}
\thankstext{e2}{{\it e-mail}:
\href{mailto:alessandro.papa@fis.unical.it}{alessandro.papa@fis.unical.it}}

\institute{European Centre for Theoretical Studies in Nuclear Physics and Related Areas (ECT*), I-38123 Villazzano, Trento, Italy\label{addr1}
\and
Fondazione Bruno Kessler (FBK),
I-38123 Povo, Trento, Italy\label{addr2}
\and
INFN-TIFPA Trento Institute of Fundamental Physics and Applications,
I-38123 Povo, Trento, Italy\label{addr3}
\and
Dipartimento di Fisica, Universit\`a della Calabria, I-87036 Arcavacata di Rende, Cosenza, Italy\label{addr4}
\and
Istituto Nazionale di Fisica Nucleare, Gruppo collegato di Cosenza, I-87036 Arcavacata di Rende, Cosenza, Italy\label{addr5}
}

\date{\today}

\maketitle


\section*{Abstract}
By making use of the hybrid collinear and high-energy factorization, where the BFKL resummation of leading and next-to-leading energy logarithms is combined with the standard description in terms of collinear parton densities, we compare predictions for Mueller-Na\-vel\-et jet rapidity and angular differential rates with data collected by CMS at $\sqrt{s} = 7$ TeV.
We provide an evidence that the study of azimuthal distributions, calculated as a Fourier sum of correlation moments and embodying the high-energy signal coming from all conformal-spin modes, permits us to overcome the well-known issues emerging in the description of Mueller-Navelet final states at \emph{natural} values of the renormalization scale.
We come out with a clear indication that the next-to-leading BFKL description of these observables at natural scales is valid when the rapidity interval between the two jets is large, and it allows us to catch the core high-energy dynamics emerging from data.
\vspace{0.30cm} \hrule
\vspace{0.30cm}
{
 \setlength{\parindent}{0pt}
 \textsc{Keywords}: \vspace{0.00cm} \\ QCD phenomenology, High-energy resummation, \\ Mueller-Navelet jets, Hunting BFKL
}
\vspace{0.75cm}


\section{Introduction}
\label{sec:intro}

The study of the dynamics behind fundamental interactions at the energy frontier of the Large Hadron Collider (LHC), as well as of new-generation machines and facilities\tcite{Chapon:2020heu,Anchordoqui:2021ghd,Feng:2022inv,Hentschinski:2022xnd,Accardi:2012qut,AbdulKhalek:2021gbh,Khalek:2022bzd,Acosta:2022ejc,AlexanderAryshev:2022pkx,Brunner:2022usy,Arbuzov:2020cqg,Abazov:2021hku,Bernardi:2022hny,Amoroso:2022eow,Celiberto:2018hdy,Klein:2020nvu,2064676,MuonCollider:2022xlm,Aime:2022flm,MuonCollider:2022ded,Begel:2022kwp,Dawson:2022zbb}, relies upon our ability of doing more and more accurate calculations by means of hi\-gher-order perturbative techniques.
Here, important challenges come from Quantum Chromodynamics (QCD), its (non)perturbative dual nature bringing to yet unresolved puzzles.
The founding pillar of QCD is the well-established \emph{collinear factorization} between hard-parton scatterings and nonperturbative parton distribution functions (PDFs) and fragmentation functions (FFs), whose validity has been corroborated by a long list of theoretical achievements and experimental evidences.
There exist kinematic regions where, however, a pure collinear, fixed-order description fails because it misses large contributions which are logarithmically enhanced in the considered phase-space corner(s). Thus, collinear factorization must be improved and supplemented by including those large terms \emph{via} all-order techniques, known as \emph{resummations}.

One of this region is the Regge-Gribov or \emph{semihard} regime\tcite{Gribov:1983ivg}, where the stringent scale hierarchy $s\gg \{Q\}^2 \gg \LQCD^2$ ($s$ is the center-of-mass energy squared, while $\{Q\}$ stands for one or a set of hard scales typical of the considered final state) leads to the rise of $\ln(s/Q^2)$ type logarithms, which enter the perturbative series with a power increasing with the order of the strong coupling, $\alpha_{s}$.
The most powerful formalism that allows us to account for these large energy logarithms to all orders is the Balitsky-Fadin-Kuraev-Lipatov (BFKL) resummation~\cite{Fadin:1975cb,Kuraev:1976ge,Kuraev:1977fs,Balitsky:1978ic}. 
It permits us to systematically catch all contributions proportional to $[\alpha_s\ln(s)]^n$, in the leading-logarithmic (LL) approximation, and of those accompanying powers of $\alpha_s[\alpha_s\ln(s)]^n$, in the next-to-leading logarithmic (NLL) approximation.

BFKL cross sections are cast as high-energy convolutions between a Green's function, which encodes the resummation of energy logarithms and does not depend on the given process, and two impact factors, portraying the fragmentation of each incoming object. The evolution of the Green's function is controlled by an integral equation, whose kernel was calculated within the next-to-leading order (NLO) in the perturbative expansion for any fixed, not increasing with $s$, momentum transfer $t$ and for any possible two-gluon colored exchange in the $t$-channel\tcite{Fadin:1998py,Ciafaloni:1998gs,Fadin:1998jv,Fadin:2000kx,Fadin:2000hu,Fadin:2004zq,Fadin:2005zj}.
Impact factors depends on the process, so that they represent the most challenging pieces of the cross section.
They are known at NLO for a limited selection of final states: (\emph{i}) quarks and gluons\tcite{Fadin:1999de,Fadin:1999df,Ciafaloni:1998kx,Ciafaloni:1998hu,Ciafaloni:2000sq}, namely the building blocks to compute (\emph{ii}) forward-jet\tcite{Bartels:2001ge,Bartels:2002yj,Caporale:2011cc,Ivanov:2012ms,Colferai:2015zfa} and (\emph{iii}) forward light-hadron\tcite{Ivanov:2012iv} impact factors, then (\emph{iv}) the impact factor for the light vector-meson electroproduction, (\emph{v}) the ($\gamma^* \to \gamma^*$) impact factor\tcite{Bartels:2000gt,Bartels:2001mv,Bartels:2002uz,Bartels:2003zi,Bartels:2004bi,Fadin:2001ap,Balitsky:2012bs}, and (\emph{vi}) the one describing the forward-Higgs production in gluon fusion in the infinite top-mass limit\tcite{Hentschinski:2020tbi,Celiberto:2022fgx}.

Suitable channels whereby to hunt for the onset of BFKL dynamics are reactions featuring the scattering of particles with small transverse sizes, such as the ones that can be investigated in lepton-antilepton collisions. Here, the absence of any initial-state hadronic activity permits us to pick the high-energy signal in a very clean way.
The growth with energy of cross sections, predicted by BFKL, was observed in total $(\gamma^* \gamma^*)$ rates~\cite{Brodsky:1998kn,Brodsky:2002ka,Caporale:2008is,Zheng:2013uja,Chirilli:2014dcb,Ivanov:2014hpa}. However, comparisons with the only available data at LEP2 were unsatisfactory, due to the low center-of-mass energies and insufficient detector accuracies.
NLL results were provided for the exclusive diffractive electroproduction of two light vector mesons~\cite{Ivanov:2005gn,Ivanov:2006gt,Enberg:2005eq} and for the photoproduction of two $\JPsi$ particles~\cite{Kwiecinski:1998sa}.
The leading-order (LO) impact factor depicting the photoproduction of forward heavy-quark pairs was recently obtained in Ref.~\cite{Celiberto:2017nyx} (see Ref.\tcite{Bolognino:2019yls} for the corresponding calculation in hadroproduction), while first results for rapidity distributions and azimuthal-angle correlations for the double heavy-qu\-ark pair photoemission were studied at LEP2 energies as well as at nominal ones of future lepton linear colliding machines~\cite{Celiberto:2017nyx,Bolognino:2019ccd}.

Notably, the high-energy resummation offers us an intriguing opportunity to access the proton structure at small-$x$ \emph{via} single-forward detections.
In particular, it provides us with a formal definition of the \emph{unintegrated gluon distribution} (UGD) in the proton, written in terms of a convolution in the transverse-momentum space\tcite{Catani:1990xk,Catani:1990eg,Catani:1993ww,Ball:2007ra,Caola:2010kv} between the BFKL Green's function and a soft, nonperturbative proton impact factor.
Extensive tests of the UGD were done through deep-inelastic-scattering structure functions\tcite{Hentschinski:2012kr,Hentschinski:2013id} and light vector-meson polarized amplitudes and cross sections at HERA\tcite{Anikin:2009bf,Anikin:2011sa,Besse:2013muy,Bolognino:2018rhb,Bolognino:2018mlw,Bolognino:2019bko,Bolognino:2019pba,Celiberto:2019slj} and, quite recently, at the Electron-Ion Collider (EIC)\tcite{Bolognino:2021niq,Bolognino:2021gjm,Bolognino:2022uty,Celiberto:2022fam,Bolognino:2022ndh}.
Further studies were done in the context of forward Drell-Yan\tcite{Motyka:2014lya,Brzeminski:2016lwh,Motyka:2016lta,Celiberto:2018muu} and vector-quarkonium\tcite{Bautista:2016xnp,Garcia:2019tne,Hentschinski:2020yfm,Goncalves:2018blz,Cepila:2017nef,Guzey:2020ntc,Jenkovszky:2021sis,Flore:2020jau,ColpaniSerri:2021bla} final states.
Starting from the information about the gluon motion inside the proton carried by the UGD, first determinations of small-$x$ improved collinear PDFs and transverse-momentum-dependent (TMD) polarized gluon TMDs were achieved in Refs.\tcite{Ball:2017otu,Bonvini:2019wxf} and\tcite{Bacchetta:2020vty,Celiberto:2021zww,Bacchetta:2021oht,Bacchetta:2021lvw,Bacchetta:2021twk,Bacchetta:2022esb,Bacchetta:2022crh,Bacchetta:2022nyv,Celiberto:2022omz}, respectively.

The weight of small-$x$ effects from BFKL in ha\-dro\-nic collisions was quantified by studying inclusive rates for the single-central emission of a Higgs boson in gluon fusion\tcite{Marzani:2008az,Caola:2011wq,Forte:2015gve}.
By making use of the Altarelli-Ball-Forte (ABF) prescription\tcite{Ball:1995vc,Ball:1997vf,Altarelli:2001ji,Altarelli:2003hk,Altarelli:2005ni} to embody small-$x$ logarithms inside the Dokshitzer-Gribov-Li\-pa\-tov-Altarelli-Parisi (DGLAP) approach,
studies in Refs.\tcite{Bonvini:2018ixe,Bonvini:2018iwt} have provided an evidence that the BFKL resummation becomes more and more relevant as the center-of-mass energy increases, up to correct the Higgs total cross section by 10\% when $\sqrt{s}$ reaches 100 TeV, \emph{i.e.} the nominal energy of the Future Circular Collider (FCC)\tcite{Mangano:2016jyj}.
Coming back to present-day opportunities, the important task of hunting for high-energy dynamics at current energies and kinematic ranges of the LHC relies on identifying a suitable class of semihard reactions that offer us $(i)$ the possibility of comparing data already collected with NLL predictions, as well as $(ii)$ a solid stability of related differential distributions under higher-order corrections.
Such a family of processes was proposed almost two decades ago, when the study of azimuthal-angle correlations between two Mueller-Navelet jets\tcite{Mueller:1986ey} emitted with high transverse momenta and large rapidity separation\footnote{Similar analyses on $\JPsi$ plus jet, Drell-Yan plus jet, and multijet tags were proposed in Refs.\tcite{Boussarie:2017oae},\tcite{Golec-Biernat:2018kem}, and\tcite{Caporale:2015vya,Caporale:2015int,Caporale:2016soq,Caporale:2016vxt,Caporale:2016xku,Celiberto:2016vhn,Caporale:2016djm,Chachamis:2016lyi,Chachamis:2016qct,Caporale:2016pqe,Caporale:2016lnh,Caporale:2016zkc,Chachamis:2017vfa,Caporale:2017jqj}, respectively.} became feasible \emph{via} the development of the \emph{hybrid} collinear and high-energy factorization\tcite{Colferai:2010wu} (see Refs.\tcite{Deak:2009xt,vanHameren:2015uia,Deak:2018obv,VanHaevermaet:2020rro,Blanco:2020akb,vanHameren:2020rqt,Guiot:2021vnp,vanHameren:2022mtk} for another formalism similar to our one).
Mueller-Navelet final states probe incoming protons at moderate $x$-values. Thus, a collinear description in terms of PDFs remains affordable.
On the other hand, however, large rapidity distances bring to $t$-channel exchanges of high transverse momenta, so that energy logarithms are enhanced.
Thus, a hybrid formalism was set, where high-energy resummed partonic hard factors are natively calculated within BFKL, and then convoluted with collinear PDFs.

A remarkable number of phenomenological studies for Mueller-Navelet jet emissions has appeared so far. A limited selection includes works in Refs.\tcite{Angioni:2011wj,Caporale:2012ih,Ducloue:2013hia,Ducloue:2013bva,Caporale:2013uva,Caporale:2014gpa,Ducloue:2015jba,Celiberto:2015yba,Celiberto:2015mpa,Caporale:2015uva,Mueller:2015ael,Celiberto:2016ygs,Celiberto:2016vva,Caporale:2018qnm,deLeon:2021ecb}. 
The first comparison of NLL predictions for Mueller-Navelet azimuthal-correlation moments with the only experimental data analyzed so far, namely the CMS ones at $\sqrt{s} = 7\mbox{ TeV}$ and for \emph{symmetric} windows of jet transverse momenta\tcite{Khachatryan:2016udy} was done almost ten years ago\tcite{Ducloue:2013hia,Ducloue:2013bva,Caporale:2014gpa}.
It led to the conclusion that the kinematic regime accessed by current data stays in between the nominal-validity regions of BFKL and DGLAP.
Then, a clear evidence was provided\tcite{Celiberto:2015yba,Celiberto:2015mpa}  that high-energy effects can be sharply singled out from the fixed-order background at the same energies and rapidity configurations adopted in Ref.\tcite{Khachatryan:2016udy} by simply imposing \emph{asymmetric} cuts for the observed transverse momenta.
In Ref.\tcite{Celiberto:2016ygs} it was pointed out that Mueller-Navelet azimuthal correlations have a very mild dependence on dynamic constraints in the
central-ra\-pi\-di\-ty region.

The calculation of the NLO correction to the forward light-hadron impact factor\tcite{Ivanov:2012iv} made a full NLL analysis of two-hadron\tcite{Celiberto:2016hae,Celiberto:2016zgb,Celiberto:2017ptm,Celiberto:2017uae,Celiberto:2017ydk} and hadron-jet\tcite{Bolognino:2018oth,Bolognino:2019cac,Bolognino:2019yqj,Celiberto:2020wpk,Celiberto:2020rxb,Celiberto:2021xpm,Celiberto:2022kxx} semihard distributions possible.
Reference\tcite{Celiberto:2020wpk} confirmed that the BFKL dynamics fairly decouples from the high-energy DGLAP pattern also when hadrons are detected (eventually, in association with jets) in asymmetric transverse-momentum windows.
The same study, however, highlighted difficulties emerging when these light objects, jets or hadrons, enter the definition of BFKL cross sections and angular distributions studied around the \emph{natural} renormalization and factorization scales suggested by the process kinematics.
They are due to resummation instabilities generated by the weight of NLL corrections, which are large and with opposite sign of pure LL contributions.
As a result, differential cross sections can easily reach negative values for large values of the rapidity interval between the detected particles. Furthermore, angular correlations written in terms of cosines of multiples of the azimuthal-angle distance, $\langle \cos (n \varphi) \rangle$, exhibit an unphysical behavior, with values larger than one both in the small and in the large rapidity range.

Different strategies were proposed to cure this issue.
The Brodsky-Lepage-Mackenzie (BLM) method \cite{Brodsky:1996sg,Brodsky:1997sd,Brodsky:1998kn,Brodsky:2002ka}, as designed in its semihard version\tcite{Caporale:2015uva}, became rapidly popular, since it permitted to partially quench these instabilities on azimuthal correlations and to marginally improve the agreement with data. However, its application was ineffective on cross sections for hadron-jet final states. Indeed, the resulting optimal energy scales turned out to be sensibly larger than the natural ones\tcite{Celiberto:2017ius,Bolognino:2018oth,Celiberto:2020wpk}, with a substantial lowering of the total cross section. Thus, any attempt at reaching the precision level was unsuccessful.

A big step forward in semihard phenomenology was taken very recently, by proposing the use of the forward-hadron NLO impact factor computed in Ref. \cite{Ivanov:2012iv} to depict emissions of heavy-flavored particles at large transverse momenta. An evidence was provided that the description of forward single-charmed ($\Lambda_c$ baryons \cite{Celiberto:2021dzy,Bolognino:2022wgl}) or single-bottomed ($B$ mesons together with $\Lambda_b$ baryons \cite{Celiberto:2021fdp}) hadrons is valid when heavy-flavor FFs\tcite{Kneesch:2007ey,Kniehl:2008zza,Kramer:2018vde,Kramer:2018rgb,Kniehl:2020szu} determined in the variable-flavor number-scheme (VFNS)\tcite{Mele:1990cw,Cacciari:1993mq} are embodied in the hybrid factorization.
However, the most striking success was the discovery that the peculiar behavior in energy of the gluon fragmentation to a heavy hadron leads to a \emph{natural stabilization} of the high-energy series, with a substantial dampening of instabilities associated to higher-order corrections. Thus, fair studies of rapidity- and angular-differential distributions became feasible at natural scales, with NLL results for heavy-flavored species being much closer to LL ones than what happens for light jets and hadrons.
A subsequent analysis on vector quarkonia\tcite{Celiberto:2022dyf} and $B_c^{(*)}$ mesons\tcite{Celiberto:2022keu} done by combining BFKL, collinear PDFs and nonrelativistic-QCD FFs \cite{Braaten:1993mp,Braaten:1993rw,Zheng:2019dfk,Zheng:2019gnb,Zheng:2021sdo}, highlighted that the natural stability is an \emph{intrinsic} property of heavy-flavor emissions. It emerges whenever a heavy-flavored bound state is tagged, independently of the \emph{Ans\"atz} made in determining or modeling its production from single-parton fragmentation.
An analogous stabilization pattern was found in partially-NLL resummed distributions sensitive to the hadroproduction of a Higgs-jet system\tcite{Celiberto:2020tmb,Celiberto:2021fjf,Celiberto:2021tky,Celiberto:2021txb}.

With the discovery of the natural stability, the hybrid factorization gained enough reliability to be employed as a powerful tool to gauge the possibility of making precision studies of high-energy QCD in ultraforward rapidity directions reachable at new-generation infrastructures. Novel studies on light me\-sons plus heavy flavor\tcite{Celiberto:2022rfj} and on charmed hadrons plus Higgs bosons \cite{Celiberto:2022zdg} have supported experimental plans of making ATLAS and the Forward-Physics-Facility (FPF) detectors\tcite{Anchordoqui:2021ghd,Feng:2022inv} work in coincidence \emph{via} a narrow-timing setup.
The main goal there has been increasing the motivation toward high-energy QCD studies, as a core element of the multifrontier activities which give life to FPF research programs.

In the present work we come back to comparing NLL results for Mueller-Navelet jet distributions calculated in hybrid factorization with data collected by CMS at 7 TeV.
First, by providing a prime analysis of systematic effects coming from scale variations around their natural or BLM-optimized values, we highlight how difficulties in the description of angular-correlation moments rise both from intrinsic instabilities inside BFKL and from the emergence of Sudakov \emph{threshold} logarithms which are genuinely discarded in our formalism.

Then, by analyzing the behavior of novel predictions for azimuthal-angle distributions, we corroborate the statement, formulated in the context of more exclusive semihard final states, that picking the whole signal coming from the azimuthal modes allows us to enhance the stability of BFKL at NLL.
Starting from this checkpoint, we build \emph{truncated} azimuthal distributions that collect the first segment of the experimental signal.
Their agreement with corresponding theory results at NLL becomes clearer and clearer as the rapidity interval between the two jets grows. Thus, we brace the message that current CMS data definitely contain strong high-energy imprints, which can be singled out by studying angular-dependent distributions.
The outcome of this work can serve as a useful guidance for forthcoming experimental analysis on Mueller-Navelet jet correlations at 13 TeV collision energies.

\section{Theoretical setup}
\label{sec:theory}

In this Section we present theoretical ingredients to build Mueller-Navelet differential distributions \emph{via} the hybrid factorization. The NLL resummed cross section in presented in Section\tref{ssec:theory}, while our selection for final-kinematics is shown in Section\tref{ssec:kinematics}. Sections\tref{ssec:BLM} and\tref{ssec:uncertainty} provide us with useful information on the BLM scale-optimization procedure and on our strategy to assess the weight of main uncertainties affecting our phenomenological analysis, respectively.

\begin{figure*}[!t]
\centering
\includegraphics[width=0.65\textwidth]{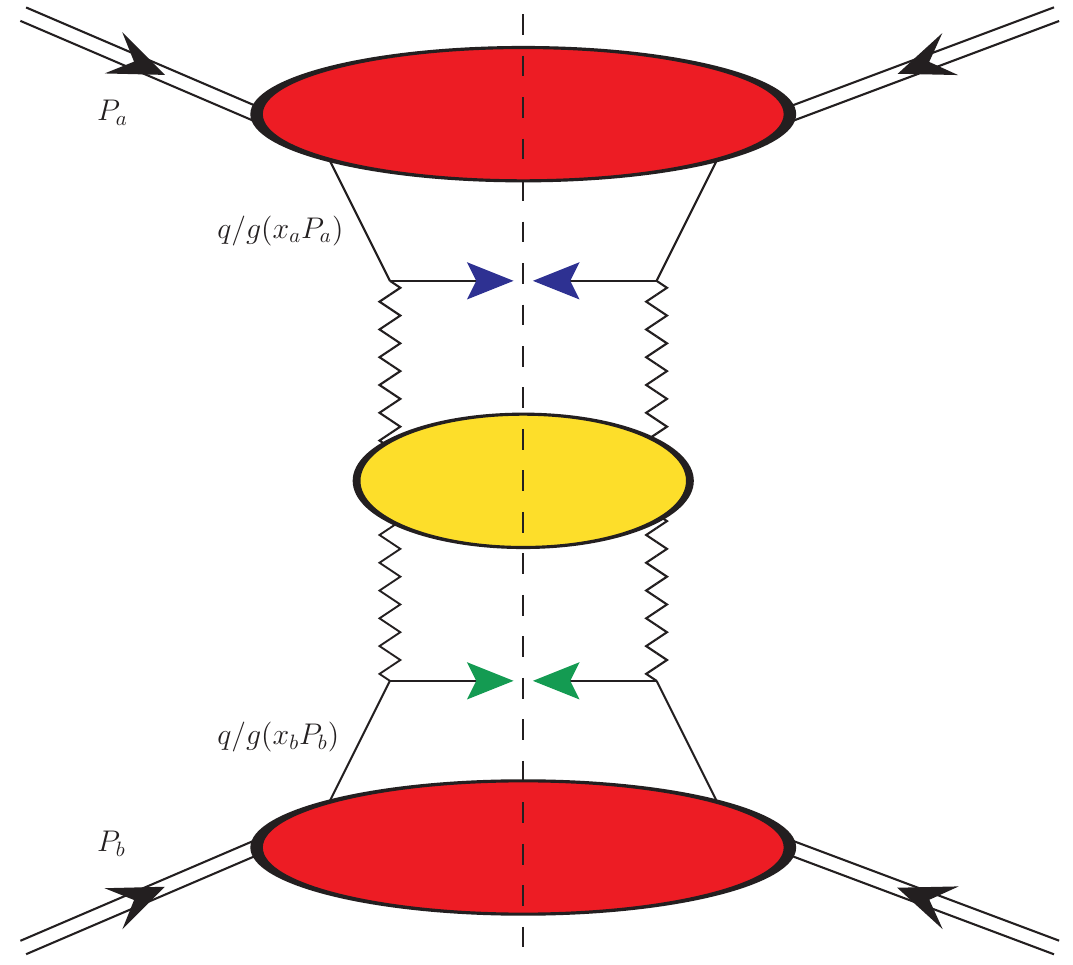}

\caption{Diagrammatic representation of the Mueller-Navelet jet hadroproduction. Red blobs denote proton collinear PDFs, while green and blue arrows denote final-state jets. The BFKL ladder, portrayed by the yellow blob, is connected to impact factors trough Reggeon (zigzag) lines. Diagrams were done \emph{via} {\tt JaxoDraw 2.0}~\cite{Binosi:2008ig}.}
\label{fig:process}
\end{figure*}

\subsection{NLL/NLO$^+$ cross section}
\label{ssec:theory}

We consider the inclusive semihard production in proton collisions of a Mueller-Navelet system (see Fig.~\ref{fig:process})
\begin{equation}
\label{eq:process}
 {\rm p}(P_a) + {\rm p}(P_b) \,\to\, {\rm jet}(q_1, y_1) + {\cal X} + {\rm jet}(q_2 , y_2) \;,
\end{equation}
The two tagged jets feature high momenta, $|\qTab| \gg \Lambda_{\rm QCD}$, and large rapidity separation, $\DY \equiv y_1 - y_2$. An undetected gluon system, ${\cal X}$, is inclusively emitted together with jets.
Our Sudakov vector basis is the one generated by incoming protons' momenta, $P_{a,b}$, with $P_{a,b}^{2} = 0$ and $({P_a} \prodscal {P_b}) = s/2$.
Thus, we decompose the final-state transverse momenta on that basis
\begin{equation}
\label{Sudakov}
 q_{1,2} = x_{1,2} P_{a,b} + \frac{\qTab^{\,2}}{x_{1,2} s} P_{b,a} + q_{{1,2 \perp}} \;,
\end{equation}
with $q_{1,2 \perp}^2 \equiv - \qTab^{\,2}$.
Working in the center-of-mass frame, one has the following relations between jet longitudinal fractions, rapidities, and transverse momenta
\begin{equation}
\label{xyp}
 x_{1,2} = \frac{|\qTab|}{\sqrt{s}} e^{\pm y_{1,2}} \;, \qquad \drv y_{1,2} = \pm \frac{d x_{1,2}}{x_{1,2}} \;,
\end{equation}
and
\begin{equation}
\label{Delta_Y}
 \qquad \DY \equiv y_1 - y_2 = \ln\frac{x_1 x_2 s}{|\qTa| |\qTb|} \;.
\end{equation}

The first step to build the high-energy resummed Mueller-Navelet cross section is the use of collinear factorization
\label{sigma_collinear}
\begin{eqnarray}
 &&\frac{\drv \sigma}{\drv x_1 \drv x_2 \drv^2 \qTa \drv^2 \qTb}
 =\sum_{\alpha,\beta=q,{\bar q},g}\int_0^1 \drv z_1 \int_0^1 \drv z_2 
 \\ \nonumber
 &&\quad\times \,
 f_\alpha\left(z_1,\mu_F\right)
 f_\beta\left(z_2,\mu_F\right)
 \frac{\drv {\hat\sigma}_{\alpha,\beta}\left(z_1 z_2 s,\mu_F\right)}
 {\drv x_1 \drv x_2 \drv^2 \qTa \drv^2 \qTb}\;,
\end{eqnarray}
where the ($\alpha, \beta$) indices run over quarks, antiquarks, and gluons, $f_{\alpha,\beta}\left(x, \mu_F \right)$ are initial-proton PDFs, $\mu_F$ is the factorization scale, $\drv \hat\sigma_{\alpha,\beta}\left(z_1 z_2 s,\mu_F \right)$ stands for the partonic-subprocess cross section, $ z_1 z_2 s$ being the squared center-of-mass energy of the partonic collision, equal to $x_1 x_2 s$ at LO.

It is possible to rewrite the cross section as a Fourier sum of the azimuthal-angle coefficients, ${\cal C}_{n \ge 0}$,
\begin{equation}
 \label{dsigma_Fourier}
 \frac{(2\pi)^2 \, \drv \sigma}{\drv y_1 \drv y_2 \drv |\qTa| \drv |\qTb| \drv \varphi_1 \drv \varphi_2} \!=\!
 \left[ {\cal C}_0 + 2 \! \sum_{n=1}^\infty \! \cos (n \varphi)\,
 {\cal C}_n \right]\, ,
\end{equation}
with $\varphi_{1,2}$ the jet azimuthal angles $\varphi \equiv \varphi_1 - \varphi_2 - \pi$.

By making use of the BFKL formalism, we come out with a consistent definition of NLL-resummed azimuthal coefficients. Working in the $\MSb$ renormalization scheme, one has (see Ref.~\cite{Caporale:2012ih} for technical details)
\begin{equation}
\label{Cn_NLLp_MSb}
 \CnNLLp \!\!\!\!=\!\! \frac{x_1 x_2}{|\qTa| |\qTb|} 
 \int_{-\infty}^{+\infty} \!\!\! \drv \nu \, e^{{\DY} \bar \alpha_s(\mu_R)
 \chi^{\rm NLO}(n,\nu)} 
 \end{equation}
\[ 
 \times \, \alpha_s^2(\mu_R) \, c_{J_1}^{\rm NLO}(n,\nu,|\qTa|, x_1)[c_{J_2}^{\rm NLO}(n,\nu,|\qTb|,x_2)]^* \;,
 \]
where $\bar \alpha_s(\mu_R) \equiv \alpha_s(\mu_R) N_c/\pi$, with $N_c$ the color number, and $\beta_0 = 11N_c/3 - 2 n_f/3$ the first coefficient of the QCD $\beta$-function.
We adopt a two-loop running-coupling setup with $\alpha_s\left(M_Z\right)=0.118$ and a dynamic number of flavors, $n_f$.
The BFKL kernel at the exponent of Eq.~\eqref{Cn_NLLp_MSb} embodies the NLL resummation of energy logarithms
\begin{eqnarray}
 \label{chi}
 \chi^{\rm NLO}(n,\nu) = \chi(n,\nu) + \bar\alpha_s \hat \chi(n,\nu) \;,
\end{eqnarray}
with $\chi(n,\nu)$ the eigenvalues of the LO BFKL kernel
\begin{eqnarray}
 \label{kernel_LO}
 \chi\left(n,\nu\right) = -2\gamma_{\rm E} - 2 \, {\rm Re} \left\{ \psi\left(\frac{1+n}{2} + i \nu \right) \right\} \, ,
\end{eqnarray}
where $\gamma_{\rm E}$ is the Euler-Mascheroni constant and $\psi(z) \equiv \Gamma^\prime
(z)/\Gamma(z)$ the logarithmic derivative of the Gamma function. Furthermore, the $\hat\chi(m,\nu)$ function in Eq.\eref{chi} represents the NLO BFKL kernel correction
\begin{eqnarray}
\label{chi_NLO}
&\hat \chi&\left(n,\nu\right) = \bar\chi(n,\nu)+\frac{\beta_0}{8 N_c}\chi(n,\nu)
\\ \nonumber &\times& 
\left\{-\chi(n,\nu)+10/3+2\ln\left[\left(\mu_R^2/(|\qTa| |\qTb|\right)\right]\right\} \;,
\end{eqnarray}
the characteristic $\bar\chi(m,\nu)$ function being calculated in Ref.~\cite{Kotikov:2000pm}.

The two expressions
\begin{equation}
\label{IFs}
c_{J_{1,2}}^{\rm NLO}(n,\nu,|\qT|,x) =
c_{J_{1,2}} +
\alpha_s(\mu_R) \, \hat c_{J_{1,2}}
\end{equation}
stand for the forward-jet NLO impact factors, calculated in the Mellin space \emph{via} the projection onto LO BFKL eigenfunctions.
The LO impact factor takes the following form:
\begin{equation}
 \label{LOJIF}
 c_J(n,\nu,|\qT|,x) = \rho_c
 |\qT|^{2i\nu-1}\,\hspace{-0.05cm}\Big[\tau_c f_g(x)
 +\hspace{-0.15cm}\sum_{\alpha=q,\bar q}\hspace{-0.10cm}f_\alpha(x)\Big] \;,
\end{equation}
with $\rho_c = 2 \sqrt{C_F/C_A}$, and $\tau_c = C_A/C_F$, where $C_F = (N_c^2-1)/(2N_c)$ and $C_A \equiv N_c$ are the Casimir factors connected to gluon emission from quark and gluon, respectively.
The formula for the NLO impact factor can be obtained by combining Eq.~(36) of~\cite{Caporale:2012ih} with Eqs.~(4.19)-(4.20) of~\cite{Colferai:2015zfa}.
It is based on calculations done in Refs.\tcite{Ivanov:2012iv,Ivanov:2012ms}, suited to numerical studies, which encode a jet algorithm calculated in the ``small-cone'' approximation (SCA)~\cite{Furman:1981kf,Aversa:1988vb} with a cone-type selection (see Ref.~\cite{Colferai:2015zfa} for further details).

Equations~(\ref{Cn_NLLp_MSb})~and~(\ref{LOJIF}) elegantly show how our hybrid collinear and high-energy factorization is realized. The cross section is factorized \emph{à la} BFKL in terms of the gluon Green's function and the impact factors. The latter ones encode collinear PDFs.
The $+$ superscript in the $\CnNLLp$ label reflects that our representation for azimuthal coefficients in Eq.\eref{Cn_NLLp_MSb} contains terms beyond the NLL accuracy generated both by the NLO exponentiated kernel and by the cross product of the NLO impact-factor corrections.
Finally, by neglecting all NLO contributions in Eq.\eref{Cn_NLLp_MSb}, we get the pure LL limit of our azimuthal coefficients
\begin{equation}
\label{Cn_LL_MSb}
\CnLL = \frac{x_1 x_2}{|\qTa| |\qTb|} 
 \int_{-\infty}^{+\infty} \drv \nu \, e^{{\DY} \bar \alpha_s(\mu_R)\chi(n,\nu)}
 \end{equation}
\[
\times \; \alpha_s^2(\mu_R) \, c_{J_1}(n,\nu,|\qTa|, x_1)[c_{J_2}(n,\nu,|\qTb|,x_2)]^* \;,
\]
which we will employ in our phenomenological analysis for comparisons with corresponding $\NLLp$ calculations.

In previous studies on Mueller-Navelet jets\tcite{Celiberto:2015yba,Celiberto:2015mpa} and hadron-jet\tcite{Celiberto:2020wpk} final states a high-energy DGLAP formula was developed to mimic the high-energy limit of a pure NLO calculation. In our jet-jet case, it can be obtained by truncating the NLL resumed azimuthal coefficients in Eq.\eref{Cn_NLLp_MSb} up to the ${\cal O}(\alpha_s^3)$ order. In this way we isolate the leading-power asymptotic signal of a pure NLO DGLAP calculation, discarding at the same time factors which are suppressed by inverse powers of $x_1 x_2 s$.
BFKL versus high-energy DGLAP analyses leads to a clear discrimination between the two approaches where asymmetric transverse-momentum configurations are adopted\tcite{Celiberto:2015yba,Celiberto:2015mpa,Celiberto:2020wpk,Celiberto:2020rxb,Celiberto:2021dzy,Celiberto:2022rfj,Celiberto:2022keu}. However, the use of our high-energy fixed order method in a theory versus experiment quest for Mueller-Navelet jets would be less adequate, since data collected by CMS are for symmetric transverse-momentum ranges only\tcite{Khachatryan:2016udy}.
Thus, instabilities emerging from higher-order calculations\tcite{Andersen:2001kta,Fontannaz:2001nq} as well as NLL energy-momentum violations NLL\tcite{Ducloue:2014koa}, which would be quenched by asymmetric cuts, prevent us to use our DGLAP expansion of resummed azimuthal coefficients.

\subsection{Final-state kinematics}
\label{ssec:kinematics}

Core ingredients to build our observables are the azimuthal coefficients integrated over rapidity and transverse momenta of the two detected jets, while their rapidity interval is kept fixed\hfill
\begin{eqnarray}
 \label{Cn_int}
 C_n &=&
     \int_{y_1^{\rm min}}^{y_1^{\rm max}} \hspace{-0.20cm} \drv y_1
 \int_{y_2^{\rm min}}^{y_2^{\rm max}} \hspace{-0.20cm} \drv y_2
 \, \,
 \delta (y_1 - y_2 - \DY)
 \\ \nonumber
 &\times&
 \int_{q_{T_1}^{\rm min}}^{q_{T_1}^{\rm max}} \hspace{-0.20cm} \drv |\qTa|
 \int_{q_{T_2}^{\rm min}}^{q_{T_2}^{\rm max}} \hspace{-0.20cm} \drv |\qTb|
 \, \,
 {\cal C}_n\left(|\qTa|, |\qTb|, y_1, y_2 \right)
 \, .
\end{eqnarray}
Here, the ${\cal C}_n$ and $C_n$ coefficients can refer to corresponding calculations taken within the NLL (Eq.~(\ref{Cn_NLLp_MSb})) or the LL (see Eq.~(\ref{Cn_LL_MSb})) accuracy.
In order to match realistic LHC kinematic cuts, we consider standard CMS configurations\tcite{Khachatryan:2016udy}. In particular, jet rapidities lies in the interval $|y_{1,2}| < 4.7$, while jet transverse momenta stay in the symmetric range 35 GeV $< |\qTab| <$ 60 GeV. The center-of-mass energy is fixed to $\sqrt{s} = 7$~TeV.

\subsection{BLM scale optimization}
\label{ssec:BLM}

The BLM method prescribes that the \emph{optimal} value for the renormalization scale, denoted as $\mu_R^{\rm BLM}$, is the one at which all the nonconformal, $\beta_0$-dependent terms entering the analytic structure of a given observable, vanish.
A specific procedure was built to remove all the $\beta_0$-dependent pieces of semihard distribution\tcite{Caporale:2015uva}, namely both the ones in the NLO BFKL kernel and in the NLO impact factors.
We refer to this method as an ``exact'' BLM approach. It was used for the first time in light-hadron correlations\tcite{Celiberto:2016hae,Celiberto:2017ptm}, and then applied to Mueller-Navelet jet phenomenology in Refs.\tcite{Celiberto:2016ygs,Celiberto:2020wpk}. Some approximated, semianalytic BLM procedures aimed at removing either the nonconformal parts of the kernel or the ones encoded in the impact factors were applied in previous studies (see, \emph{e.g.}, Refs.\tcite{Ducloue:2013hia,Ducloue:2013bva,Caporale:2014gpa}).
As a result, $\mu_R^{\rm BLM}$ turns out to be dependent on $s$ and therefore on $\DY$. The BLM procedure can in principle be applied in any renormalization scheme. In Refs.\tcite{Brodsky:1998kn,Brodsky:2002ka} it was suggested that, in the case of BFKL calculations,
the BLM should be applied after first transitioning to the MOM scheme\tcite{Barbieri:1979be,PhysRevD.20.1420,PhysRevLett.42.1435}. This prescription is also known as BFKLP, after the Authors of Refs.\tcite{Brodsky:1998kn,Brodsky:2002ka}.
Therefore, our first step consists in making a finite transformation from the $\MSb$ scheme to the MOM one.
The expression for the strong coupling in the MOM scheme, $\alpha_s^{\rm{MOM}}$, can be obtained by inverting the following relation
\begin{equation}
\label{as_MOM}
 \as^{\MSb} = \as^{\rm MOM} \left( 1 +  \frac{\tau^{\beta} + \tau^{\rm c}}{\pi} \as^{\rm MOM} \right) \;,
\end{equation}
with
\begin{eqnarray*}
\label{T_bc}
\tau^\beta &=& - \left( \frac{1}{2} + \frac{\cal{I}}{3} \right) \beta_0 \; ,
\\
\tau^{\rm c} &=& \frac{C_A}{8}\left[ \frac{17}{2}{\cal I} +\frac{3}{2}\left({\cal I}-1\right)\xi 
 + 
 \left(1-\frac{{\cal I}}{3}\right)\xi^2-\frac{1}{6}\xi^3 \right] \; ,
\end{eqnarray*}
where ${\cal I}=-2\int_0^1\drv \gamma \frac{\ln \gamma}{\gamma^2-\gamma+1} \simeq 2.3439$, with the gauge parameter $\xi$ fixed to zero in the following.
Then, to find the BLM scale of a given azimuthal coefficient, $C_n$, one has to solve the following integral equation
\begin{equation}
\label{Cn_beta_int}
  C_n^{[\beta]} =\!\! 
  \int \drv \Phi_{1,2}(y_{1,2}, |\qTab|, \DY) \,
  \, {\cal C}_n^{[\beta]}  = 0 \, ,
\end{equation}
with $\drv \Phi_{1,2}(y_{1,2}, |\qTab|, \DY)$ the final-state differential phase space (see Section\tref{ssec:kinematics}), and
\begin{eqnarray}
\label{Cn_beta}
 {\cal C}^{[\beta]}_n
 &\propto&
 \int^{\infty}_{-\infty} \drv\nu\,e^{\DY \bar \alpha^{\rm MOM}_s(\mu^{\rm BLM}_R)\chi(n,\nu)}
\\ \nonumber &\times& 
 c_{J_1}(n,\nu,|\qTa|,x_1)[c_{J_2}(n,\nu,|\qTb|,x_2)]^*
\\ \nonumber &\times& 
\left[{\cal F}(\nu) + \bar \alpha^{\rm MOM}_s(\mu^{\rm BLM}_R) \frac{\DY}{2} \chi(n,\nu) \tilde{{\cal F}}(n,\nu) \right] \, ,
\end{eqnarray}
where
\begin{eqnarray}
\label{F_nu}
\tilde{{\cal F}}(n,\nu) &=& {\cal F}(\nu) - \frac{\chi(n,\nu)}{2} \, ,
\\
{\cal F}(\nu) &=& - \frac{1 + 4 \, {\cal I}}{3} + 2 \ln \frac{\mu^{\rm BLM}_R}{\mu_N} \, .
\end{eqnarray}
It is convenient to introduce the ratio between the BLM scale and the \emph{natural} one, suggested by the process kinematics
$\mu_N \equiv \sqrt{|\qTa| |\qTb|}$, so that $ C_{\mu}^{\rm BLM} \equiv \mu_R^{\rm BLM}/\mu_N$. We look for $C_{\mu} \equiv C_{\mu}^{\rm BLM}$ values which solve Eq.~(\ref{Cn_beta_int}). We set $\mu_F = \mu_R$ everywhere.
Finally, the BLM scale value is plugged into expressions for the integrated coefficients, and we obtain the following $\NLLp$ expression in the MOM scheme
\begin{eqnarray}
\label{Cn_int_NLLp_BLM_MOM}
 &&\hspace{-0.25cm}C_{n, \, \rm{BLM}\mbox{-}\rm{MOM}}^\NLLp = \!\!
 \int \! \drv \Phi(y_{1,2}, |\qTab|, \DY) 
 \frac{x_1 x_2}{|\qTa| |\qTb|} 
 \\ \nonumber
 &&\hspace{-0.25cm}\times
 \int_{-\infty}^{+\infty} \!\! \drv \nu
 \left[\alpha^{\rm MOM}_s (\mu^{\rm BLM}_R)\right]^2
 e^{\DY \bar \alpha^{\rm MOM}_s(\mu^{\rm BLM}_R)\tilde{\chi}^{\rm NLO}(n,\nu)}
 \\ \nonumber
 &&\hspace{-0.25cm}\times
 \left\{ \tilde{c}_{J_1}^{\rm NLO}(n,\nu,|\qTa|, x_1)[\tilde{c}_{J_2}^{\rm NLO}(n,\nu,|\qTb|,x_2)]^* \right. \\
 \nonumber && \hspace{-0.25cm} \left. + \ \alpha^{\rm MOM}_s(\mu^{\rm BLM}_R) \frac{2 \tau^{\rm c}}{\pi} \right\} \;,
\end{eqnarray}
with
\begin{eqnarray}
 \label{chi_tilde}
 \tilde{\chi}^{\rm NLO}(n,\nu) &=& \chi(n,\nu) \\ \nonumber
 &+&\bar \alpha^{\rm MOM}_s(\mu^{\rm BLM}_R) \left[\bar \chi(n,\nu) +\frac{\tau^{\rm c}}{3} \chi(n,\nu)\right] 
\end{eqnarray}
\begin{equation}
\label{IFs_sub}
\tilde{c}_{J_{1,2}}^{\rm NLO}(n,\nu,|\qT|,x) =
c_{J_{1,2}} +
\bar\alpha_s^{\rm MOM}(\mu^{\rm BLM}_R) \, \bar c_{J_{1,2}} \;,
\end{equation}
where $c_{J_{1,2}}$ are the LO impact factors (Eq.\eref{LOJIF}), while $\bar c_{{1,2}}(n,\nu,|\qTab|,x_{1,2})$ denotes the NLO impact-factor corrections after the removal of $\beta_0$-dependent factors, the being universal factors proportional to LO impact factors. We have
\begin{equation}
\label{IF_NLO_sub}
\bar c_{J_{1,2}} = \hat c_{J_{1,2}} + \frac{\beta_0}{4 N_c} \left[ \mp \, i \frac{\drv}{\drv \nu} \!+\! \left( \ln \mu_R^2 + \frac{5}{3} \right) \right] c_{J_{1,2}} \, .
\end{equation}
In order to compare data with resummed predictions both at natural scales and at BLM-optimized ones in the same renormalization scheme, we need the corresponding formula of Eq.\eref{Cn_int_NLLp_BLM_MOM} in the $\MSb$ scheme.
It can be obtained by making the two following substitutions in Eq.\eref{Cn_int_NLLp_BLM_MOM}
\[
 \alpha_s^{\rm MOM}(\mu^{\rm BLM}_R) \,\to\, \alpha_s^{\MSb}(\mu^{\rm BLM}_R) \;,
\]
\begin{eqnarray}
 \label{MOM_2_MSb}
 \tau^{\rm c} \,\to\, - \tau^\beta \;.
\end{eqnarray}
More in particular, we replace the analytic expression of the $\alpha_s$ in the MOM scheme, which was obtained by inverting Eq.\eref{as_MOM}, with the corresponding $\MSb$ one, meanwhile the value of $\mu_R$ is left unchanged.

\subsection{Uncertainty estimation}
\label{ssec:uncertainty}

To get an accurate description of our process we need to identify the main potential sources of uncertainty.
Assessing the sensitivity of our observables on variations of renormalization and factorization scales is a strategy, largely used in perturbative calculations, to guess the weight of higher-order corrections.
In a recent study on semihard emissions of light and heavy hadrons\tcite{Celiberto:2022rfj}, it came out that the sensitivity on the energy-scale variation and the collinear-FF set choice gives a major contribution to the global uncertainty of cross sections. Since the theoretical description of our Mueller-Navelet final states does not rely on any fragmentation mechanism, we can reasonably assume that scale variation is the main source of uncertainty.
We will gauge the effect of simultaneously varying $\mu_R$ and $\mu_F$ around their \emph{natural} values or their BLM \emph{optimal} ones, in the range 1/2 to two.
The $C_{\mu}$ parameter entering the figures in Section\tref{sec:results} stands for the ratio $C_\mu = \mu_{R}/\mu_N$ or $C_\mu = \mu_{R}^{\rm BLM}/\mu_N$, respectively. In both cases, we will set $\mu_F$ equal to $\mu_R$.

Another potentially relevant uncertainty could co\-me from proton collinear PDFs. However, recent analyses on high-energy distributions highlighted that the selection of different PDF parametrizations as well as of different members inside the same set produces no relevant impact (see Refs.\tcite{Bolognino:2018oth,Celiberto:2020wpk,Celiberto:2021fdp,Celiberto:2022rfj}). For this reason, we will employ the central member of just one PDF set, namely the novel {\tt NNPDF4.0} one\tcite{NNPDF:2021uiq,NNPDF:2021njg}. It was extracted \emph{via} global fits and through the \emph{replica} method originally proposed in Ref.\tcite{Forte:2002fg} in the context of neural-network techniques (for the sake of completeness, we refer to Ref.\tcite{Ball:2021dab} for a detailed study on ambiguities rising from \emph{correlations} between different PDFs sets). In order to be consistent with our $\NLLp$ treatment, we will use the {\tt NNPDF4.0} NLO determinations at NLO, as provided by the {\tt LHAPDF} interface~\cite{Buckley:2014ana}.
Two further potential uncertainty sources could come from $(i)$ the \emph{collinear improvement} of the NLO BFKL kernel\tcite{Salam:1998tj,Ciafaloni:2003rd,Ciafaloni:2003ek,Ciafaloni:2000cb,Ciafaloni:1999yw,Ciafaloni:1998iv,SabioVera:2005tiv}, which prescribes the inclusion of terms generated by renormalization group (RG) to impose a compatibility with the DGLAP equation in the collinear limit, and from $(ii)$ the change of renormalization scheme.
Point $(i)$ was addressed in Ref.\tcite{Celiberto:2022rfj}, and the outcome is that the effect of the collinear improvement on cross sections stays fairly inside error bands generated by energy-scale variations, and it is even less relevant in the case of azimuthal distributions.
The same study provided us with a guess of $(ii)$ the upper limit of the impact of passing from $\MSb$ to MOM renormalization scheme. It emerged that MOM predictions for cross sections are systematically larger than $\MSb$ ones, but still inside the error band generated by scale variation. We stress, however, that a consistent MOM analysis would rely on MOM-evolved PDFs, not available so far. Therefore, the complete effect of changing the renormalization scheme in semihard reactions still needs to be quantified.
In view of these considerations, we will produce uncertainty bands for our predictions by combining the effect of $\mu_R$ and $\mu_F$ variation together with the numeric error coming from the final-state multidimensional integration (Eq.\eref{Cn_int}). The former will be included for the first time in the context for Mueller-Navelet jet studies, and it is the most relevant one. The latter will be constantly kept below $1\%$ by integration routines.

\begin{figure*}[!t]
\centering

\includegraphics[scale=0.43,clip]{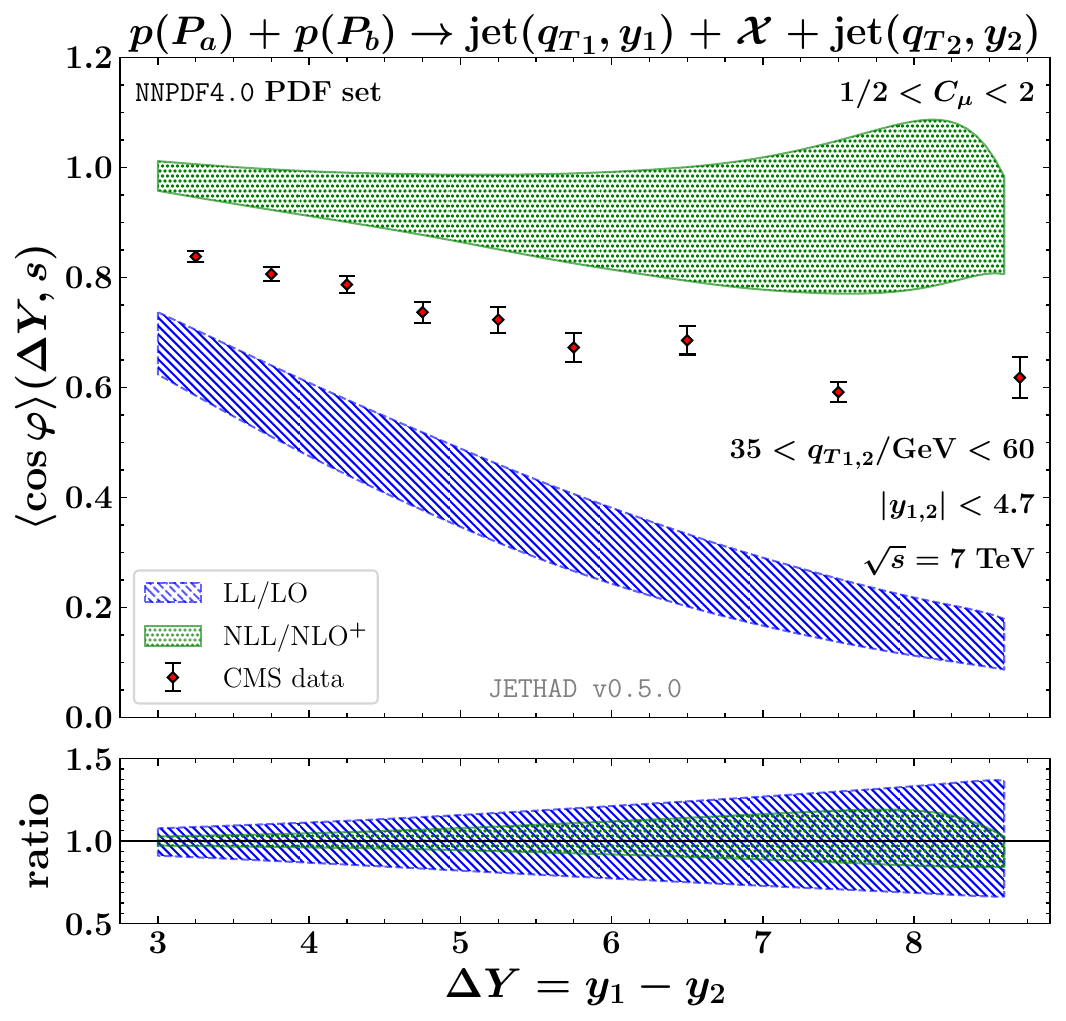}
\includegraphics[scale=0.43,clip]{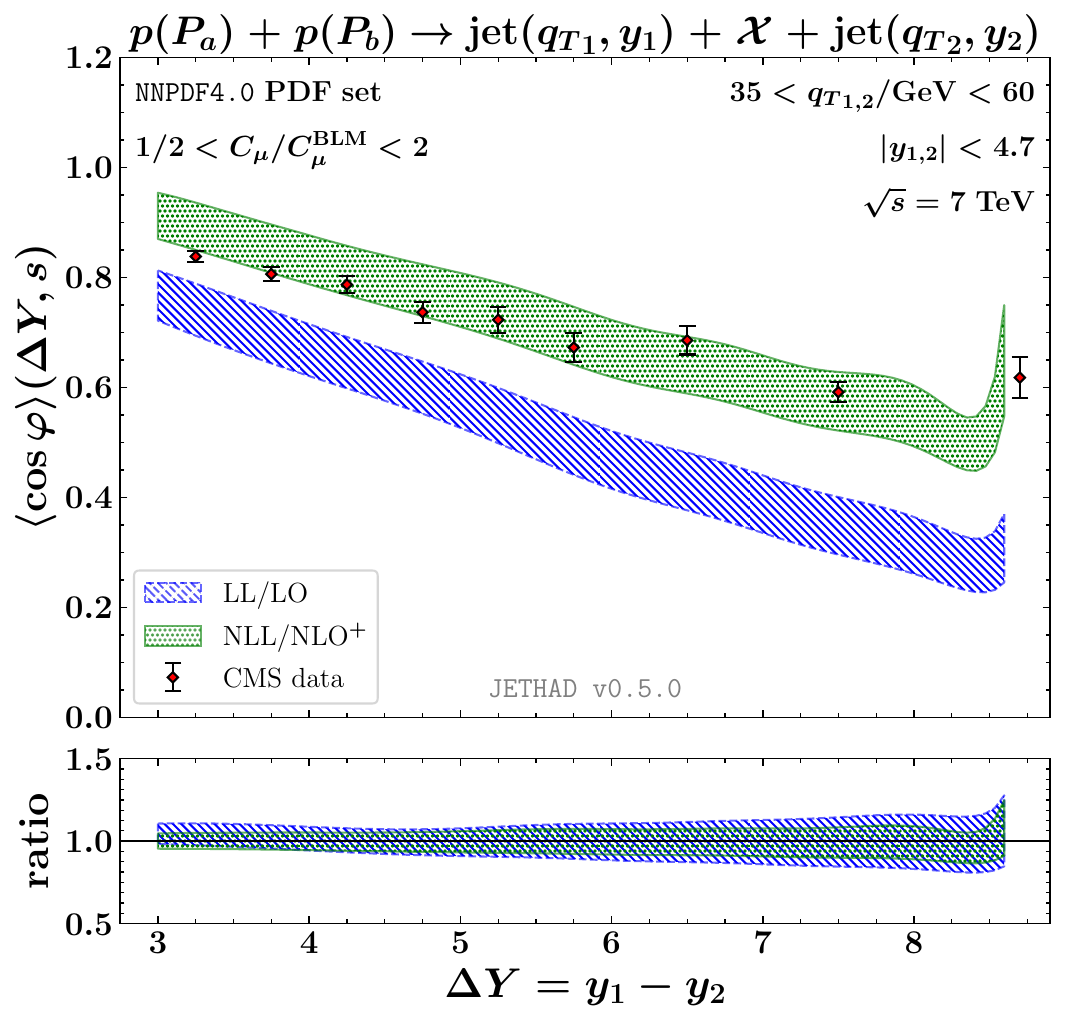}

\includegraphics[scale=0.43,clip]{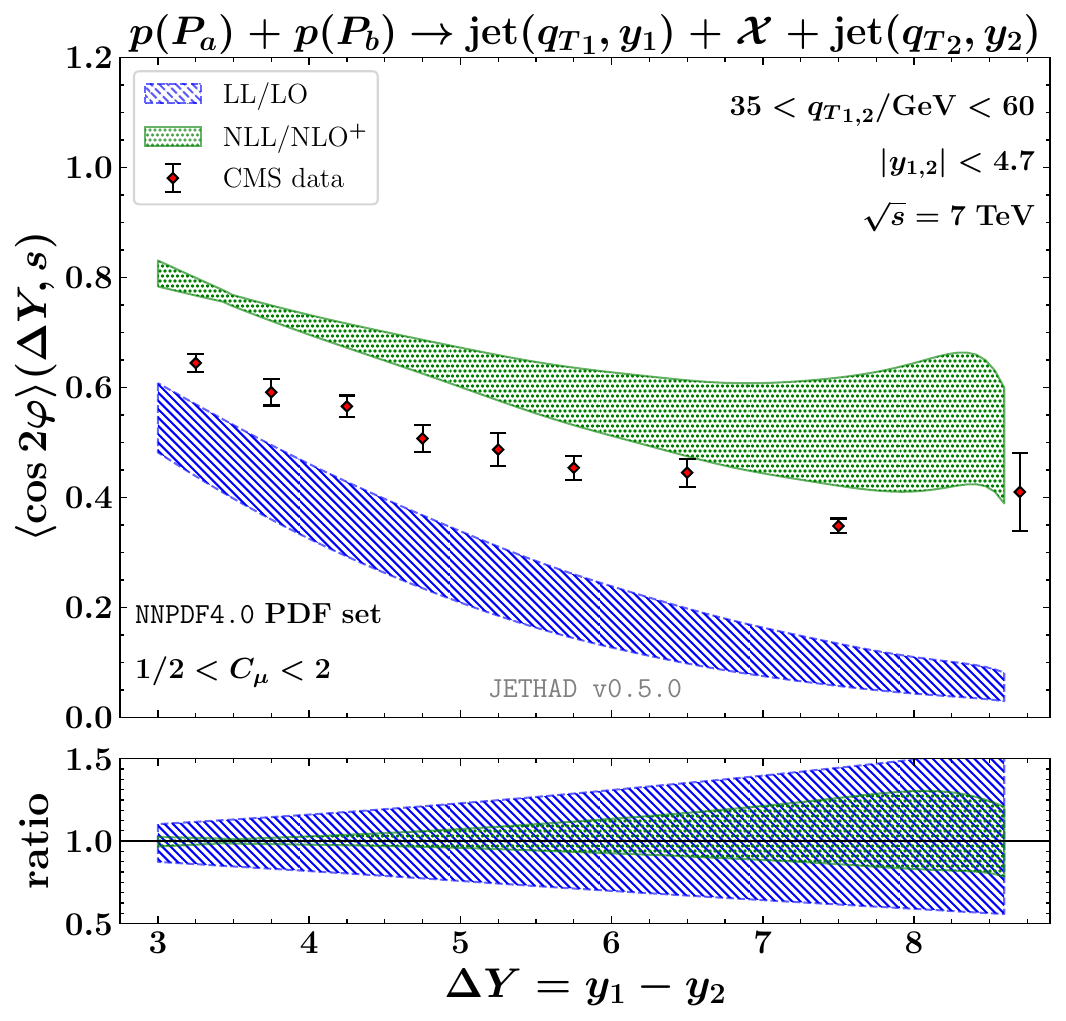}
\includegraphics[scale=0.43,clip]{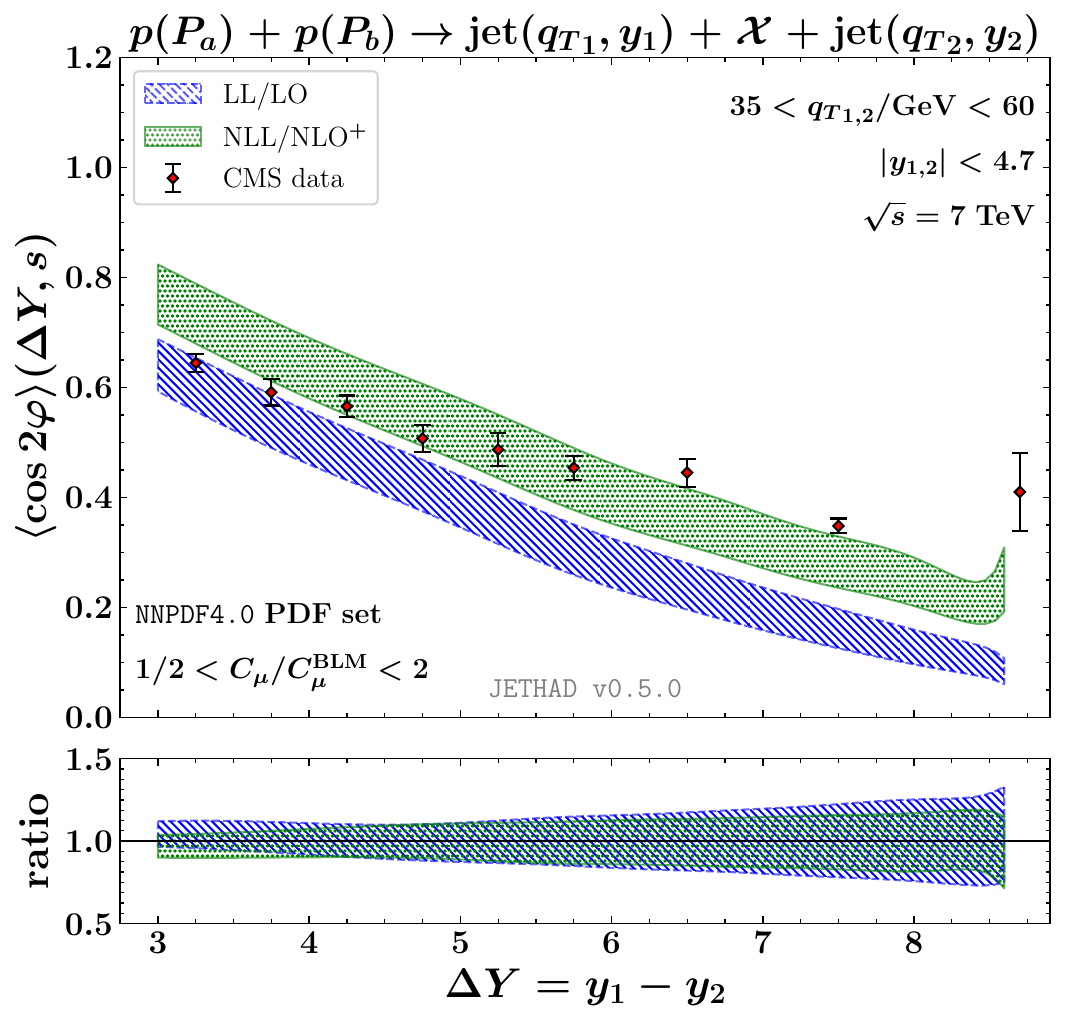}

\caption{Azimuthal-correlation moments, $R_{10} \equiv \langle \cos \varphi \rangle$ and $R_{20} \equiv \langle \cos 2 \varphi \rangle$, as functions of $\DY$ and for $\sqrt{s} = 7$ TeV, calculated at natural scales (left) or after BLM optimization (right) and compared with CMS experimental data. Text boxes inside panels exhibit final-state kinematic cuts. Uncertainty bands embody the effect of energy-scale variations. Ancillary panels below primary plots show reduced ratios, namely divided by their central values, $C_\mu = 1$ (left) or $C_\mu / C_\mu^{\rm BLM} = 1$ (right).}
\label{fig:Rn0}
\end{figure*}

\begin{figure*}[!t]
\centering

\includegraphics[scale=0.43,clip]{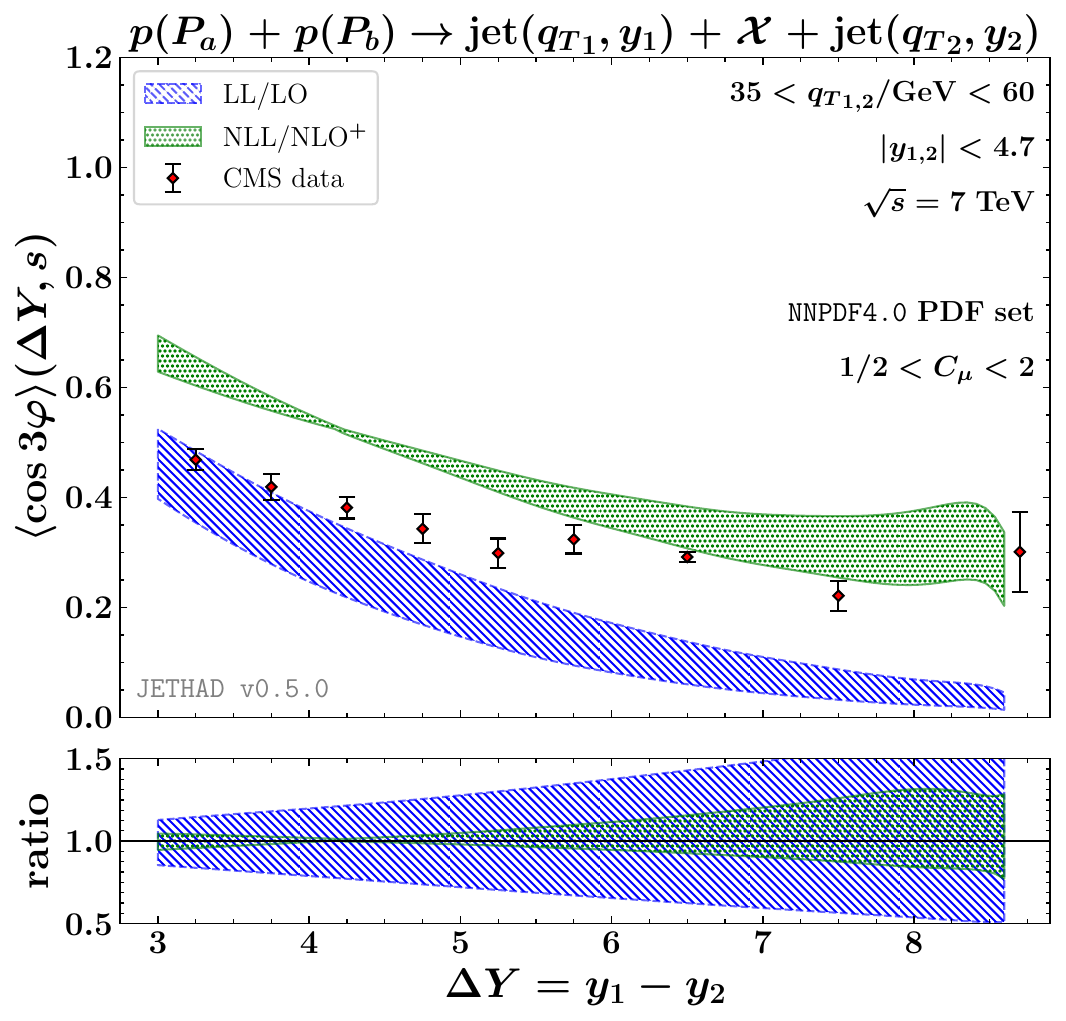}
\includegraphics[scale=0.43,clip]{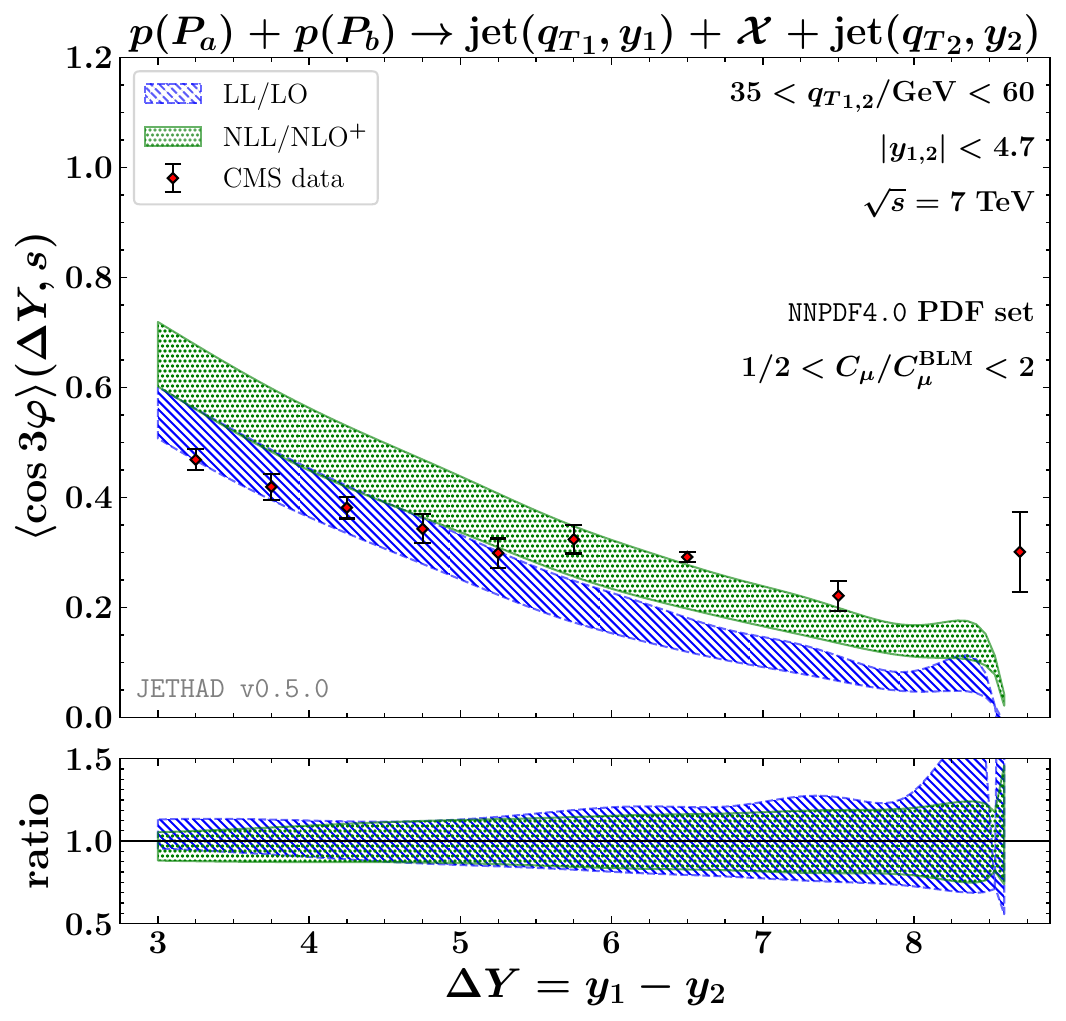}

\caption{Azimuthal-correlation moment, $R_{30} \equiv \langle \cos 3 \varphi \rangle$, as functions of $\DY$ and for $\sqrt{s} = 7$ TeV, calculated at natural scales (left) or after BLM optimization (right) and compared with CMS experimental data. Text boxes inside panels exhibit final-state kinematic cuts. Uncertainty bands embody the effect of energy-scale variations. Ancillary panels below primary plots show reduced ratios, namely divided by their central values, $C_\mu = 1$ (left) or $C_\mu / C_\mu^{\rm BLM} = 1$ (right).}
\label{fig:R30}
\end{figure*}


\section{Phenomenological analysis}
\label{sec:results}

In this Section we present results of our phenomenological study. Section\tref{ssec:Rnm} contains a comparison with CMS data of our high-energy predictions for azimuthal-correlation moments.
Section\tref{ssec:Rn0_distribution} brings an inspection of the high-energy signal rising from all azimuthal modes.
In Section\tref{ssec:phi} we provide evidence of a stabilization pattern emerging from azimuthal distributions.
In Section\tref{ssec:hunting_data} we present a strategy to compare truncated azimuthal distributions with the same CMS data for azimuthal correlations, together with an indication that these data collected at $\sqrt{s} = 7$ TeV encode clear high-energy signals.
All predictions were obtained by making use of {\tt JETHAD}, a hybrid and multimodular interface aimed at the calculation, management, and processing of observables defined by the hands of distinct formalisms\tcite{Celiberto:2020wpk,Celiberto:2022rfj}.

\subsection{Azimuthal-correlation moments}
\label{ssec:Rnm}

\begin{figure*}[!t]
\centering

\includegraphics[scale=0.43,clip]{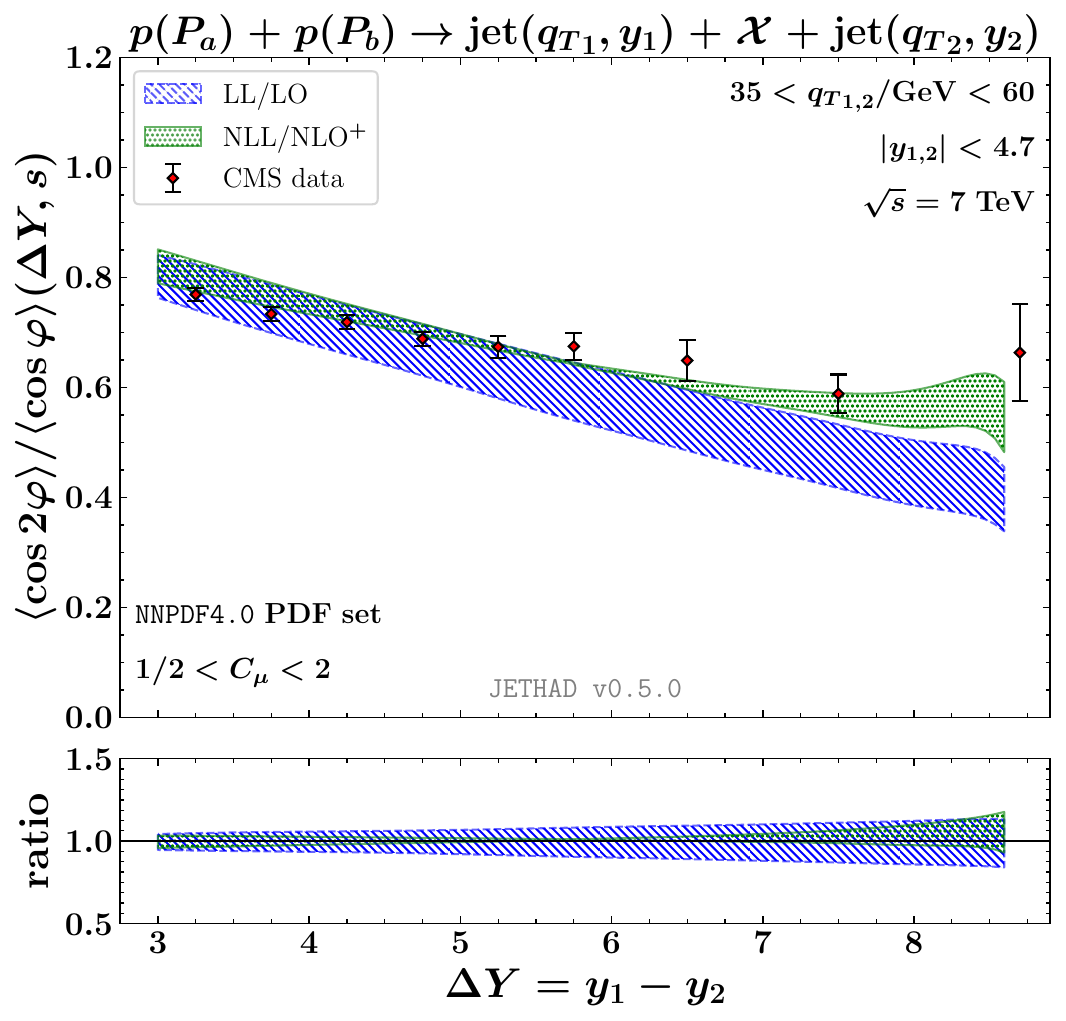}
\includegraphics[scale=0.43,clip]{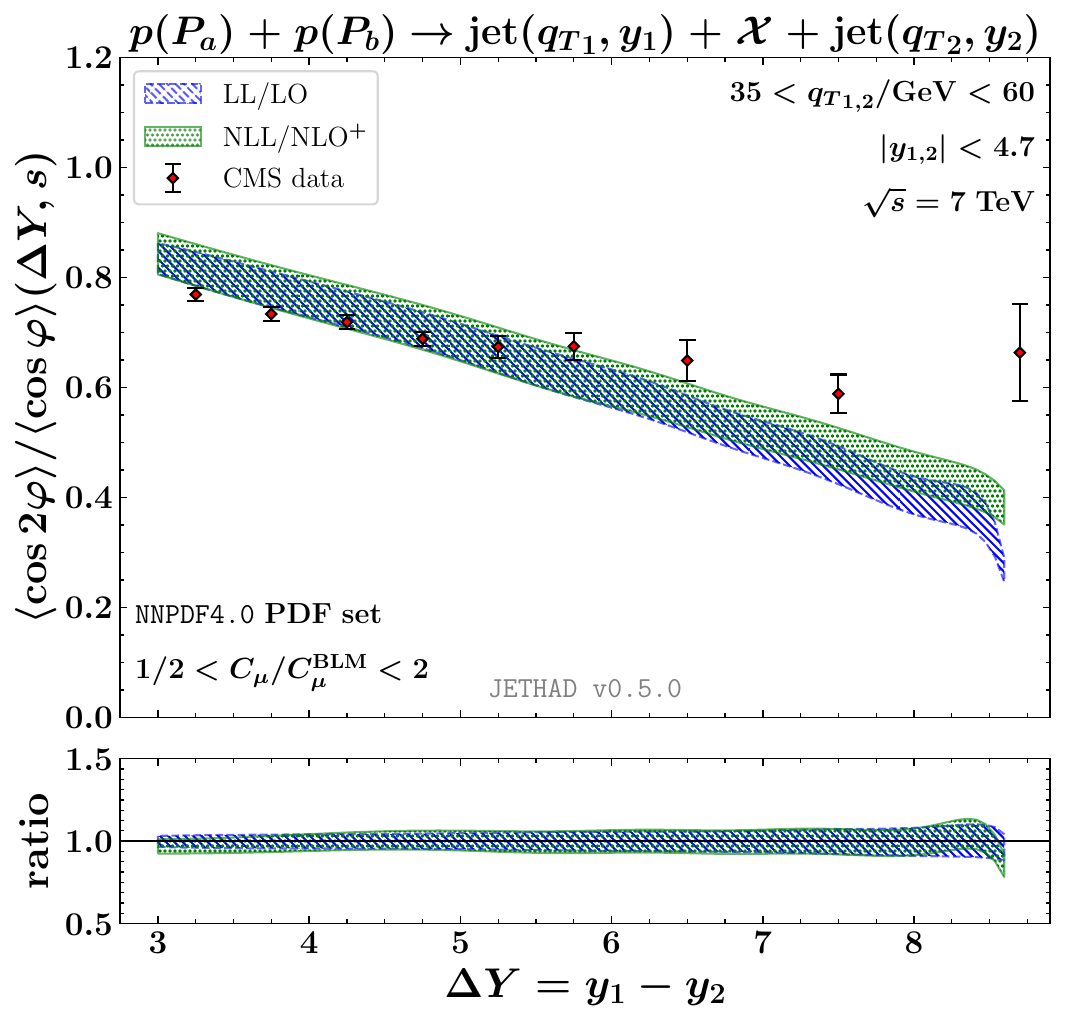}

\includegraphics[scale=0.43,clip]{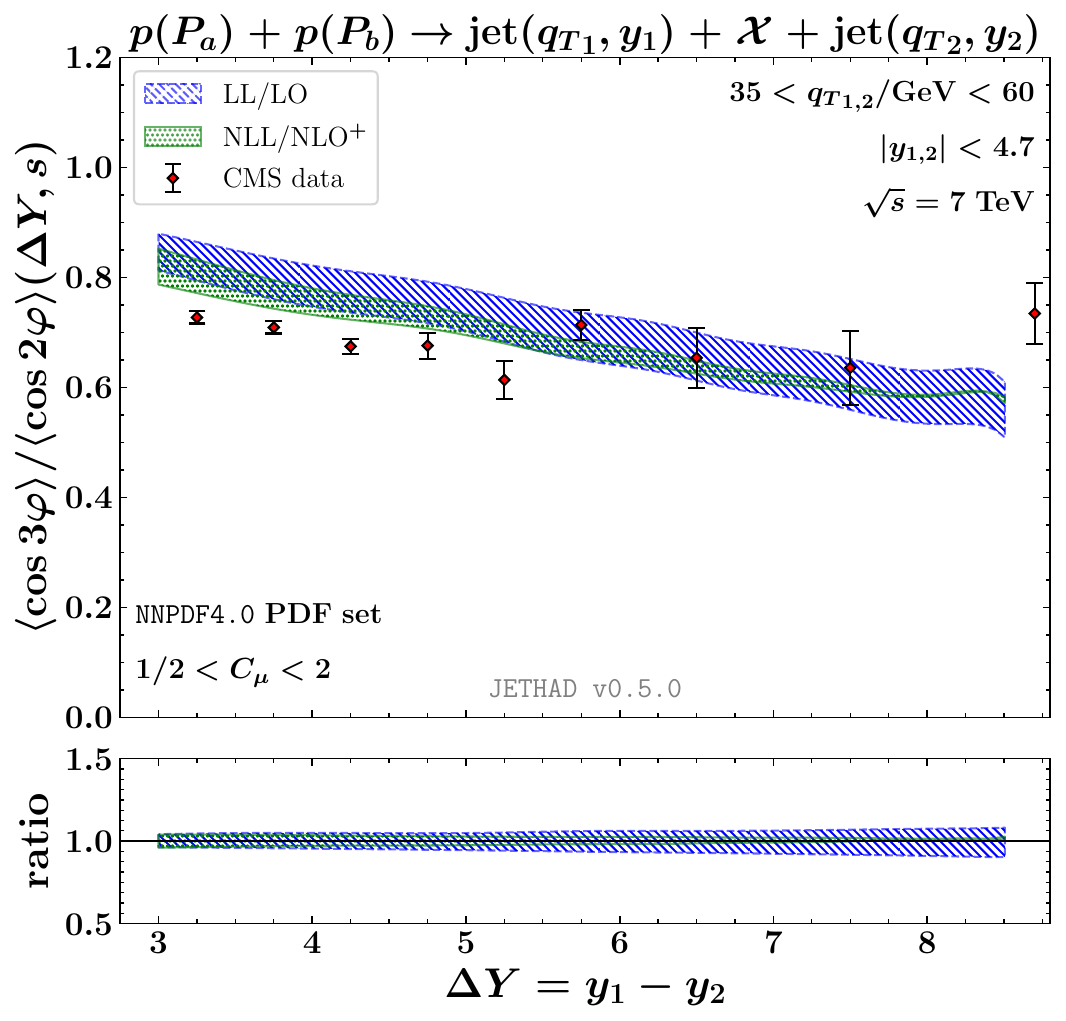}
\includegraphics[scale=0.43,clip]{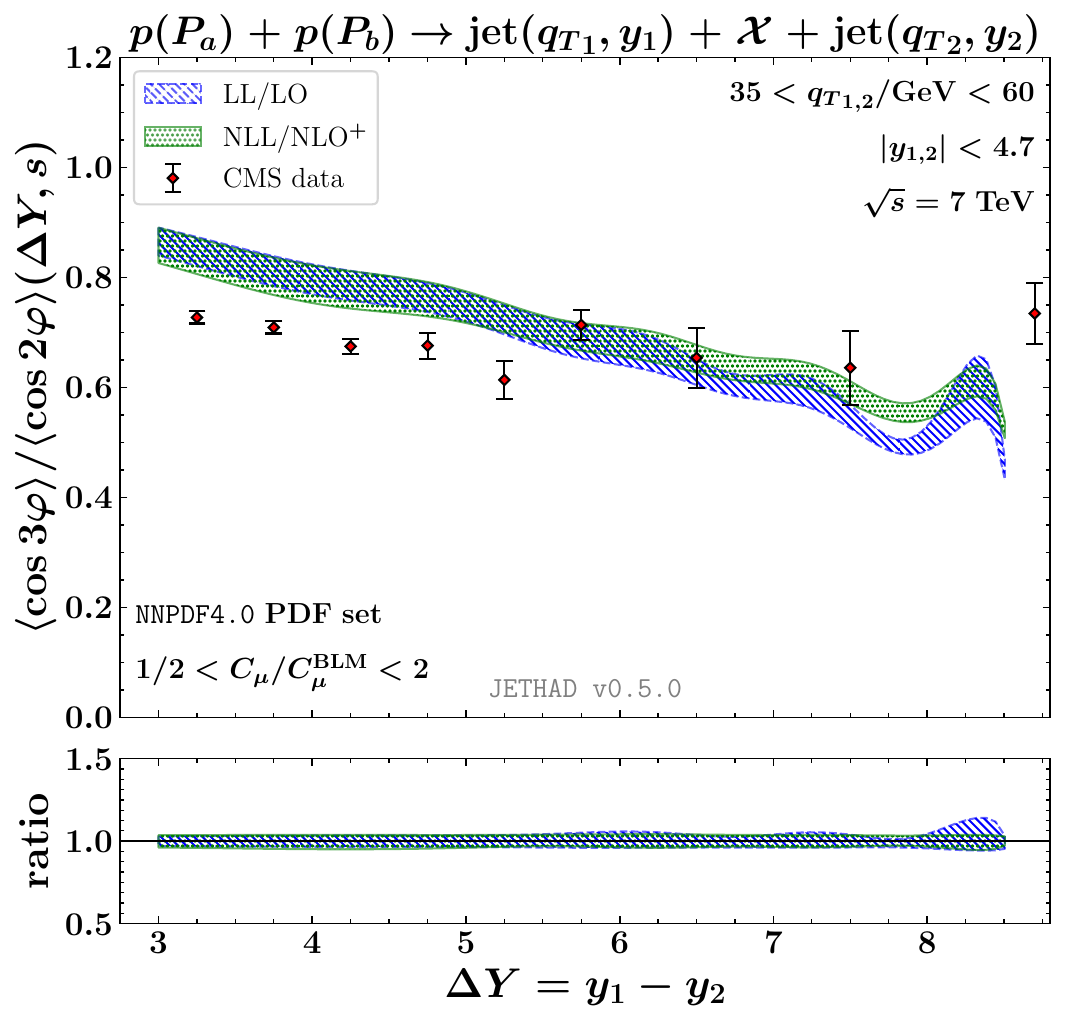}

\caption{Ratios of azimuthal-correlation moments, $R_{nm} \equiv \langle \cos n \varphi \rangle / \equiv \langle \cos m \varphi \rangle$, as functions of $\DY$ and for $\sqrt{s} = 7$ TeV, calculated at natural scales (left) or after BLM optimization (right) and compared with CMS experimental data. Text boxes inside panels exhibit final-state kinematic cuts. Uncertainty bands embody the effect of energy-scale variations. Ancillary panels below primary plots show reduced ratios, namely divided by their central values, $C_\mu = 1$ (left) or $C_\mu / C_\mu^{\rm BLM} = 1$ (right).}
\label{fig:Rnm}
\end{figure*}

Starting from the integrated azimuthal coefficients of Eq.\eref{Cn_int}, we study their ratios, $R_{nm} \equiv C_n/C_m$.
The physical interpretation of $R_{n0}$ ratios is immediate. Indeed they represent the correlation moments of the cross section, $\langle \cos n \varphi \rangle$, while the ones with $n,m > 0$ are ratios of correlations, $\langle \cos n \varphi \rangle / \langle \cos m \varphi \rangle$, proposed for the first time in Refs.~\cite{Vera:2006un,Vera:2007kn}.
In Figs.\tref{fig:Rn0} and\tref{fig:R30} we compare with CMS data the $\DY$-shape of $\LL$ $\NLLp$ $R_{n0}$ ratios, with $n=1,2,3$, at natural scales (left panels) and after applying the BLM optimization (right panels).
The overall behavior emerging here is a decreasing pattern of our predictions with $\DY$. This is an expected feature, which is shared also by data. Indeed, when the rapidity interval between the two jets grows, the phase space opens up and the weight of undetected gluons forming the ${\cal X}$ system of Eq.\eref{eq:process} becomes more and more relevant, as predicted by BFKL. Thus, the decorrelation of the two jets on the azimuthal plane becomes stronger and stronger. A pure leading-logarithmic treatment clearly overestimates such a decorrelation, with all $\LL$ predictions (blue bands) staying below data. Conversely, considering next-to-leading corrections produces a recorrelation effect. This is due to the fact that NLL BFKL terms generally have an opposite sign with respect to LL ones.\footnote{This feature holds for the BFKL Green's function and for the forward-jet NLO impact factor. In other NLO impact factors the situation could change. As an example, the $C_{gg}$ coefficient in the forward light-hadron NLO impact factor contains corrections with the same sign of LL contributions, see Refs.\tcite{Celiberto:2017ptm,Celiberto:2020wpk}.}
As already pointed out\tcite{Ducloue:2013hia,Ducloue:2013bva,Caporale:2014gpa}, the NLL recorrelation pattern is too strong when azimuthal ratios are studied at natural scales (left panels). In particular, we note that $\NLLp$ results (green bands) stays well above data. Furthermore, their distance between corresponding $\LL$ predictions is large, and their width grows with $\DY$. This represents a clear manifestation of instability of the BFKL series. Indeed, if the series were stable, the two bands would at least partially overlap and their width would shrink when the resummation accuracy increases.
Further signs of instability at natural energy scales rise from the $\NLLp$ prediction for the $R_{10}$ correlation, whose uncertainty band always contains values larger than one (left upper panel), and from negative values of all the $R_{nm}$ ratios, not shown in our plots, got when $\DY$ is larger than 8.5.
When the BLM procedure is at work (right panels), the situation improves. Here, $\NLLp$ and $\LL$ have bands closer to each other and similar in width. However, none of them is able to catch data in the full $\DY$-range, the $\NLLp$ overlapping with experimental points in the intermediate $\DY$-range.
The last observation deserves an important discussion on the interplay between the high-energy resummation and other approaches.
When $\DY$ is small, say $\DY \gtrsim 3 \div 4$, BFKL is pushed toward its limit of applicability, since the phase space opened up for secondary gluon emissions is very limited. In this region, a pure DGLAP-based approach turns out to be valid.
When $\DY$ is large, say $\DY > 8$, we enter the so-called \emph{threshold} region, namely where the energy of the
Mueller-Navelet system approaches the value of the center-of-mass energy. 
Here, collinear PDFs are probed at longitudinal fractions
to the endpoints of their definition, where they become affected by significant scaling violations and uncertainties.
Moreover, since our observables are sounded at the edges of their phase space, Sudakov effects coming from emissions of soft and collinear gluons become more and more relevant and they must be accounted for by an appropriate resummation.
Different strategies to resum threshold logarithms in rapidity-inclusive rates have been developed so far\tcite{Sterman:1986aj,Catani:1989ne,Catani:1996yz,Bonciani:2003nt,deFlorian:2005fzc,Ahrens:2009cxz,deFlorian:2012yg,Forte:2021wxe}.
A procedure to combine together both small-$x$ effects from BFKL and large-$x$ threshold ones was set up in the context of central-inclusive Higgs-boson production\tcite{Ball:2013bra,Bonvini:2018ixe,Bonvini:2018iwt}.
A similar result for Mueller-Navelet jets is not yet available. Here, the major difficulty emerges from the presence of a fixed rapidity interval between the two jets, which calls for an extension of methods mentioned above to the case of rapidity-differential rates. A consistent method to exactly resum the two species of logarithms (energy and threshold) and remove the associated double counting both in the NLL Green's function and in the NLO impact factor is a demanding task, which clearly goes beyond the scope of the present study. We limited ourselves to performing preliminary and simpler tests on gauging the impact of effectively including threshold effects by replacing the {\tt NNPDF4.0} PDF set with the large-$x$ {\tt NNPDF3.0lx} one \cite{Bonvini:2015ira}. We discovered that this change produces no visible effects on semihard observables. This is in line with the statement that threshold effects are more relevant in off-shell coefficients functions (namely impact factors) than in collinear parton densities\tcite{Bonvini:2018iwt}.
As already mentioned, $R_{nm}$ ratios of correlations were  proposed as possibly more favorable observables in the search for a clearer signals of high-energy effects\tcite{Vera:2006un,Vera:2007kn}. This comes out from the observation that eliminating the collinear poles \emph{via} RG-im\-pro\-ve\-ment techniques has a sizable effect on the asymptotic intercept of the $\varphi$-averaged cross section, $C_0$, while the intercepts corresponding to $C_{n>0}$ coefficients are hardly affected.
This translates in a stronger convergence of the latters with respect to $C_0$.
Thus, investigating observables not sensitive to the $n=0$ conformal-spin mode should be more promising.
In Fig.\tref{fig:Rnm} we present the $\DY$-behavior of $R_{21}$ and $R_{32}$ ratios of azimuthal correlations. We clearly observe that, although being very close to each other and partially nested, neither $\LL$ nor $\NLLp$ bands taken at natural scales (left panels) are compatible with CMS data in the full $\DY$-range. 
At variance with $R_{n0}$ correlations, the situation worsens when the BLM optimization is employed (right panels).
Even if they exhibit more stable patterns when passing from a pure leading to the next-to-leading accuracy, $R_{nm}$ ratios fail to complete the theory versus experiment quest.
This brings us to the guess that the high-energy signal encoded in data could be caught by other observables that genuinely embody the full high-energy signal coming both from $C_0$ and the higher-order modes (see Section\tref{ssec:Rn0_distribution}).

\subsection{High-energy signals from azimuthal modes}
\label{ssec:Rn0_distribution}

\begin{figure*}[!t]
\centering

\includegraphics[scale=0.43,clip]{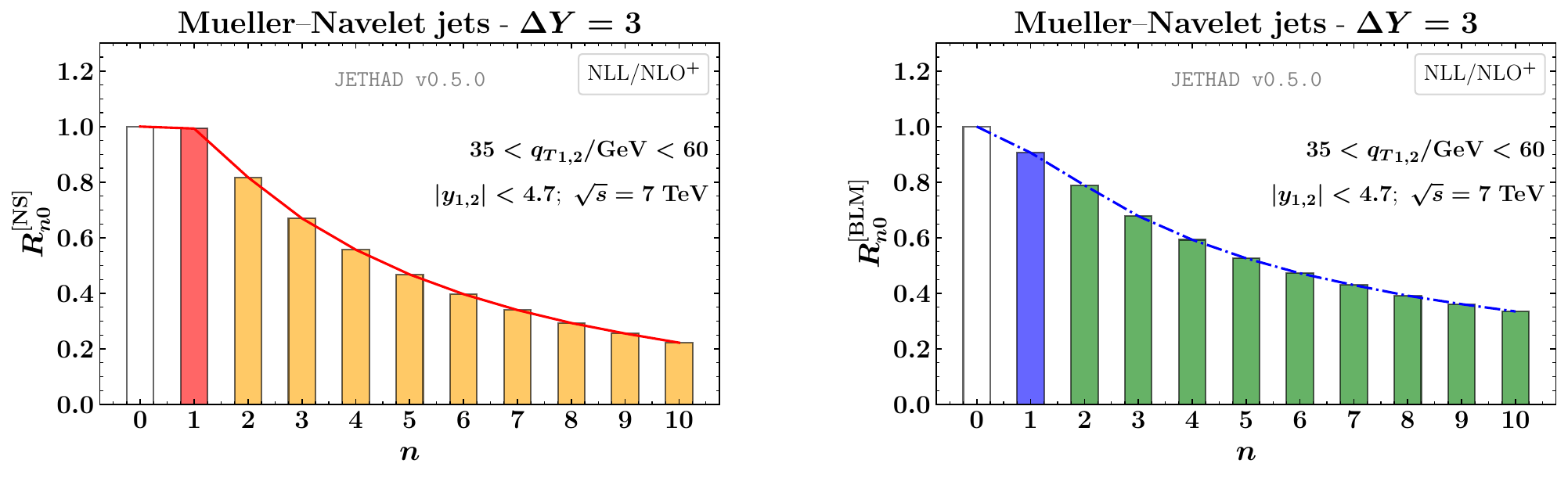}
\vskip.50cm
\includegraphics[scale=0.43,clip]{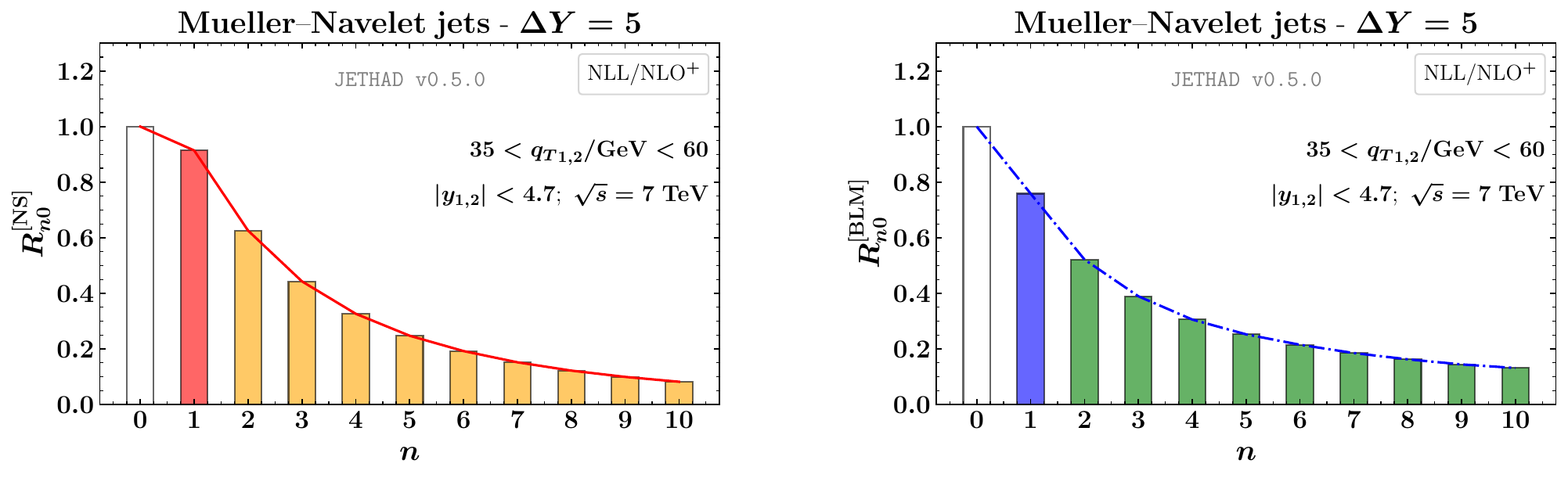}
\vskip.50cm
\includegraphics[scale=0.43,clip]{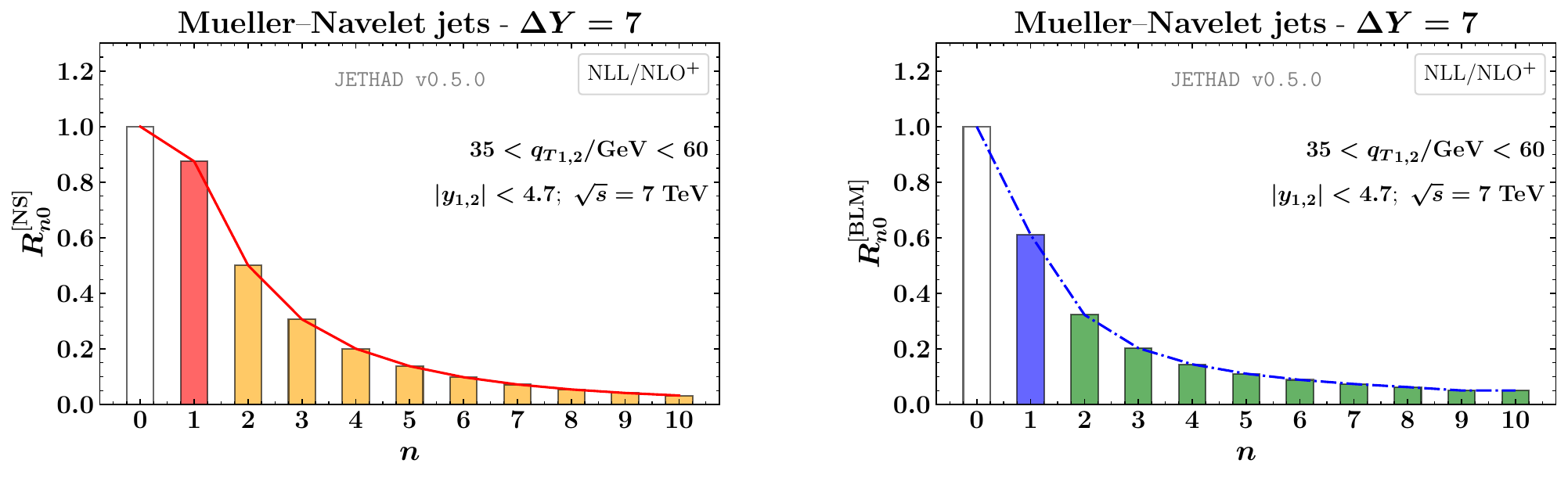}

\caption{NLL/NLO$^+$ distributions of moments, $R_{n0}$, for distinct values of $\DY$ and for $\sqrt{s} = 7$ TeV, at natural scales (left) or after BLM optimization (right). Text boxes inside panels exhibit final-state kinematic cuts.}
\label{fig:Rn0_distribution}
\end{figure*}

The main outcome of the previous Section was the need for identifying observables sensitive to the BFKL dynamics, different from the azimuthal-correlation moments and their ratios. In particular, these novel observables should be defined as functions of all the azimuthal coefficients $C_n$, and not just of $C_0$ or of a given $R_{nm}$ ratio.
To prove the robustness of this requirement, we present in Fig.\tref{fig:Rn0_distribution} the distribution of $\NLLp$ $R_{n0}$ moments, with $n$ ranging from one to 10, organized in bar charts produced at fixed values of the rapidity interval, $\DY = 3, 5, 7$. Left (right) panels are for $R_{n0}^{\rm [NS]}$ ($R_{n0}^{\rm [BLM]}$) predictions at natural (BLM) energy scales. 
To emphasize the weight of the first angular-dependent coefficient, $C_1$, over $C_0$, bars associated to $R_{10}$ are given with a different color with respect to higher moments: red versus orange at natural scales, blue versus green at BLM ones.
For the sake of brevity, uncertainty bands connected to scale variations around their natural or BLM values are not shown.
A remarkable, twofold pattern emerges from the inspection of the bar charts.
On one hand, lower $R_{n0}$ moments, say the ones with $n \le 5$, are larger at natural scales when compared with corresponding ones obtained with BLM. This is particularly true for $R_{10}^{\rm [NS]}$, which almost saturates one when $\DY = 3$, and it decreases for larger $\DY$-values, although staying well above 0.9. Conversely, $R_{10}^{\rm [BLM]}$ roughly goes from 0.9 down to 0.6 when $\DY$ grows.
On the other hand, higher $R_{n0}$ moments, say the ones with $5 < n \le 10$, exhibit an opposite behavior, with $R_{n0}^{\rm [NS]}$ being constantly lower than $R_{10}^{\rm [BLM]}$.
The origin of such a duplex pattern has to be sought by analyzing the net effect of the BLM method.
From an operational viewpoint, applying BLM leads to an expansion of scale values, which can become 10 times (or even more) larger than natural ones (see Ref.\tcite{Celiberto:2020wpk}, Fig.~4).
The $\chi(n,\nu)$ function entering the NLO kernel (see Eqs.\eref{kernel_LO} and\eref{chi_NLO}) gives a major contribution to the exponential factors in Eqs.\eref{Cn_NLLp_MSb} and\eref{Cn_int_NLLp_BLM_MOM} for lower values of $n$. In the very low-$(n,\nu)$ range $\chi(n,\nu)$ is positive. In this case, the larger is $\mu_R$, the smaller is the running coupling and thus, the exponential term. This explains why BLM scales bring to a reduction of low-$n$ correlations.
Conversely, in the large-$n$ and large-$\nu$" range $\chi(n,\nu)$ is negative. Thus, the larger is $\mu_R$, the larger is the exponential term.
This is why large-$n$ correlations are smaller at natural scales than at BLM ones.
The overall indication guessed from results of Fig.\tref{fig:Rn0_distribution} is that $(i)$ higher azimuthal modes are particularly important at natural scales and they should be encoded in the definition of BFKL-sensitive distributions, and $(ii)$ the dynamic hierarchy between $R_{n0}^{\rm [NS]}$ and $R_{n0}^{\rm [BLM]}$ as $n$ varies could balance in these novel observables. This would translate in a stabilization pattern when passing from natural to BLM scales.

\subsection{Stabilization pattern from azimuthal distributions}
\label{ssec:phi}

\begin{figure*}[!t]
\centering

\includegraphics[scale=0.43,clip]{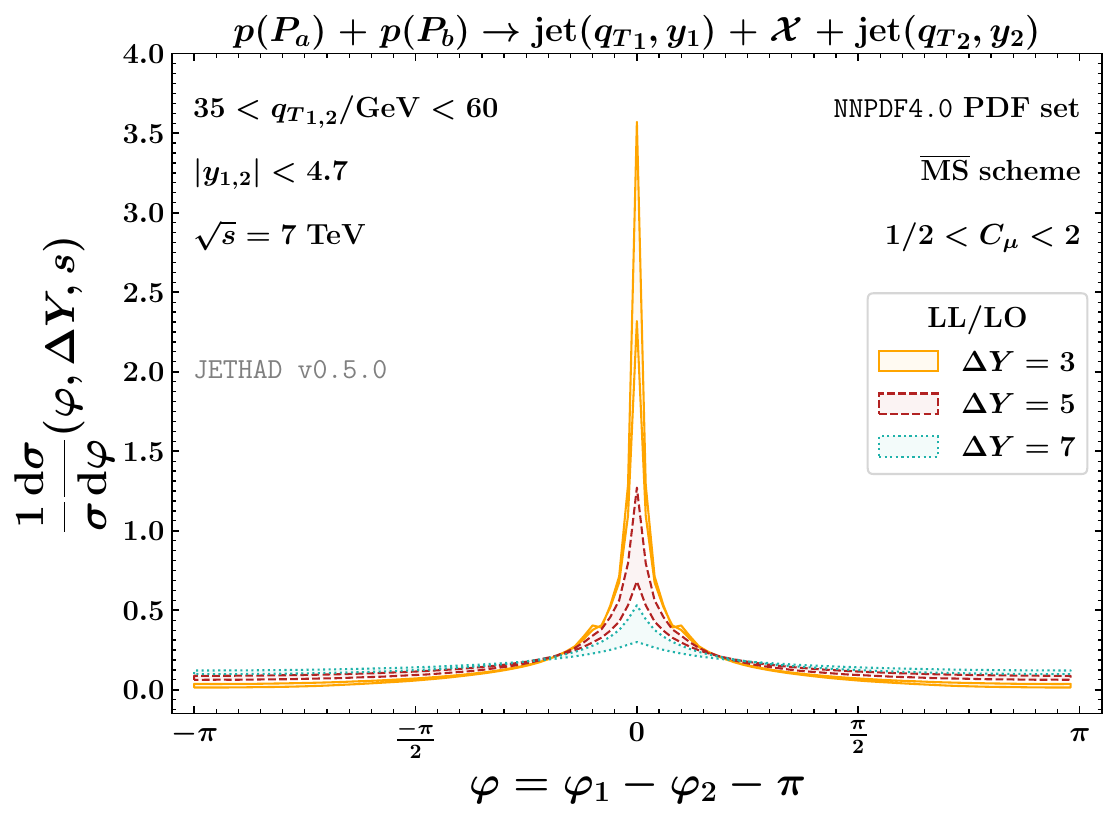}
\includegraphics[scale=0.43,clip]{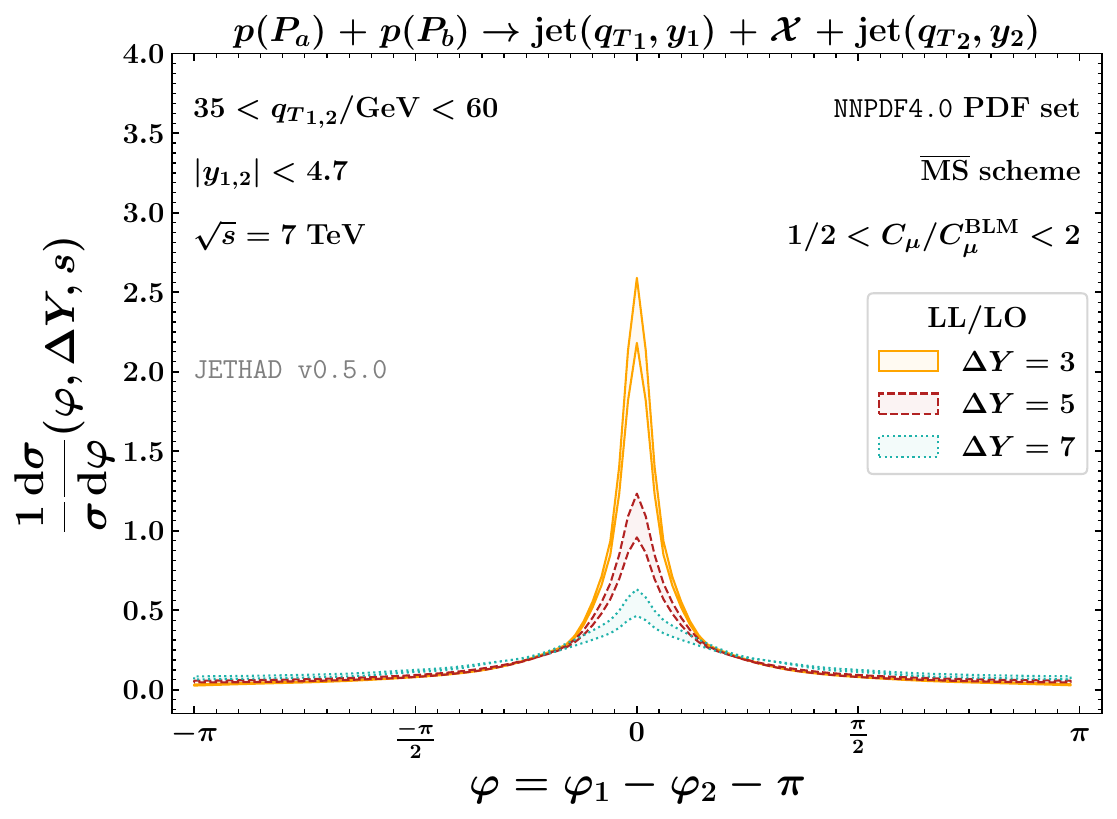}

\includegraphics[scale=0.43,clip]{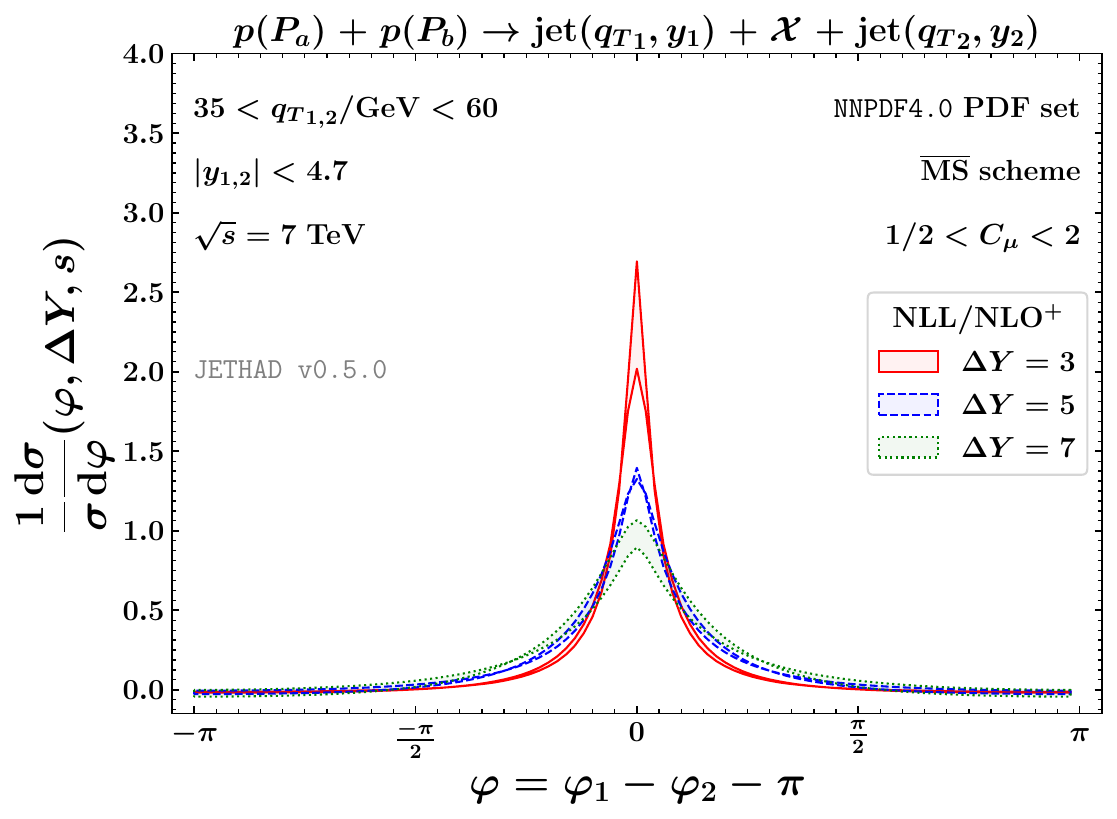}
\includegraphics[scale=0.43,clip]{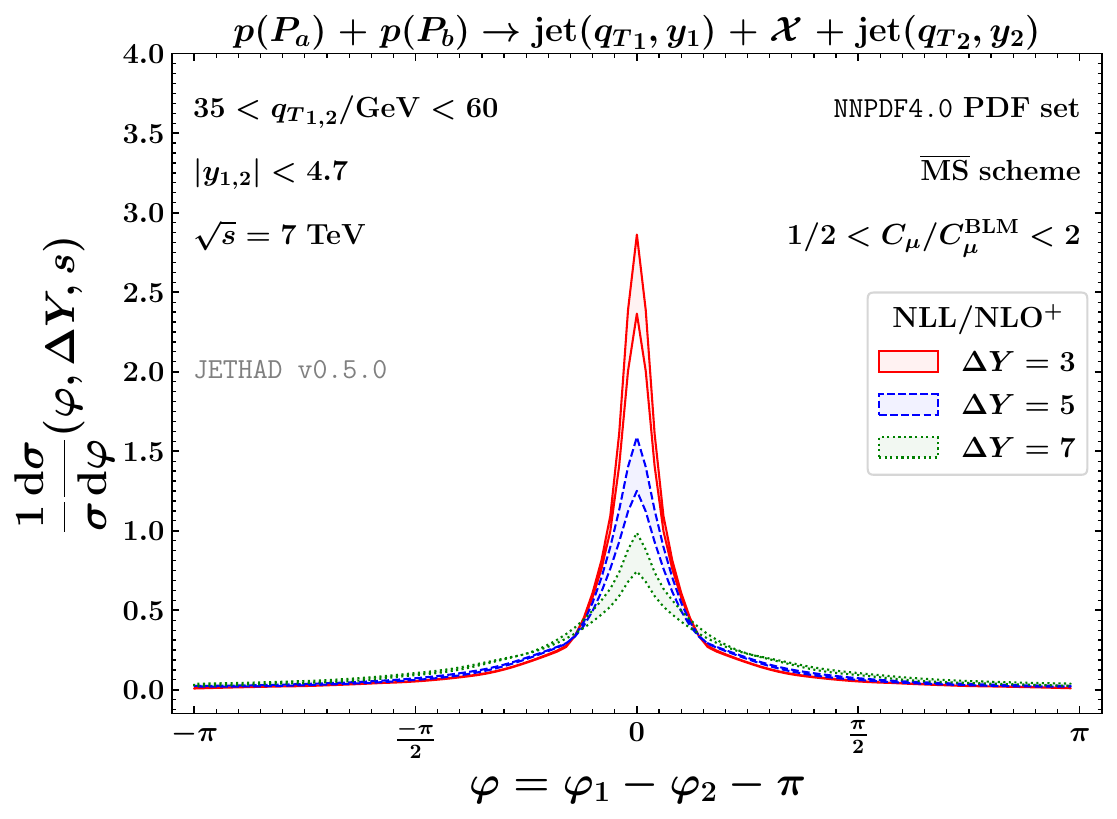}

\caption{LL/LO (upper) and NLL/NLO$^+$ (lower) azimuthal distributions for distinct values of $\DY$ and for $\sqrt{s} = 7$ TeV, at natural scales (left) or after BLM optimization (right). Text boxes inside panels exhibit final-state kinematic cuts. Uncertainty bands embody the effect of energy-scale variations.}
\label{fig:phi}
\end{figure*}

A straightforward way to probe the high-energy spectrum of our process in terms of all the azimuthal modes is considering azimuthal-angle distributions, defined as the following multiplicities
\begin{eqnarray}
\label{dsigma_dphi}
 \frac{1}{\sigma}
 \frac{\drv \sigma (\varphi, \DY, s)}{\drv \varphi} &=& \frac{1}{2 \pi} \left\{ 1 + 2 \sum_{n = 1}^{\infty} \cos(n \varphi) \langle \cos(n \varphi) \rangle \right\}
 \nonumber \\
 &\equiv& \frac{1}{2 \pi} \left\{ 1 + 2 \sum_{n =1 }^{\infty} \cos(n \varphi) R_{n0} \right\} \; .
\end{eqnarray}
As shown in Eq.\eref{dsigma_dphi}, they are built as a Fourier sum of all the $R_{n0} \equiv \langle \cos(n \varphi) \rangle$ correlations and thus, in terms of $C_0$ and the higher azimuthal coefficients, $C_{n>0}$.
These $\varphi$-dependent observables were first employed for Mueller-Navelet jets in pioneering LL analyses\tcite{Marquet:2007xx}, then used to evaluate the effect of the collinear improvement\tcite{Vera:2007kn}, to perform a first NLL versus experiment study at 7 TeV LHC\tcite{Ducloue:2013hia}, and finally to access hadron-jet correlations\tcite{Celiberto:2020wpk}.
Their study turned out to be novel in the more general context of semihard reactions, where a \emph{natural stabilization} pattern was recently discovered when Higgs bosons\tcite{Celiberto:2020tmb,Celiberto:2022zdg}, heavy jets\tcite{Bolognino:2021mrc,Bolognino:2021hxx}, singly\tcite{Celiberto:2021dzy,Celiberto:2021fdp,Celiberto:2022rfj} or doubly heavy-flavored hadrons\tcite{Celiberto:2022dyf,Celiberto:2022kza,Celiberto:2022keu} are inclusively produced in forward-rapidity directions at the LHC.
Besides encoding the full high-energy azimuthal signal and the emergence of the natural stability, relevant from a theoretical perspective, measuring azimuthal distributions is particularly advantageous from an experimental point of view.
Indeed, since detector acceptances cannot cover the entire ($2 \pi$) azimuthal-angle range, confronting theory with data for a $\varphi$-differential distribution is much easier than for a standard $R_{nm}$ ratio.
A potential issue on the computational side could rise from the infinite sum over $n$ in Eq.\eref{dsigma_dphi}, which must be necessarily truncated by the machine to a numerical cutoff, $\nu_{\rm [num]}^{\rm cut}$.
We found a fair numerical convergence for $\nu_{\rm [num]}^{\rm cut} = 50$.
In Fig.\tref{fig:phi} we show $\LL$ (upper) and $\NLLp$ (lower) predictions for Mueller-Navelet azimuthal distributions at natural (left) and BLM (right) scales, calculated at $\DY=3,5,7$.
A first glance at left plots fairly confirms the possibility of studying our distributions at natural scales, the instability emerging in $R_{nm}$ ratios (see Section\tref{ssec:Rnm}) being absent here.
Then, a clear evidence of high-energy dynamics rises from the presented analysis.
The general trend consists in the presence of a definite peak when $\varphi = 0$. At this value, the two jets are produced back-to-back.
In all the cases the peak height substantially lowers when $\DY$ increases, whereas the width of the distribution slightly broadens. 

\begin{figure*}[!t]
\centering

\includegraphics[scale=0.41,clip]{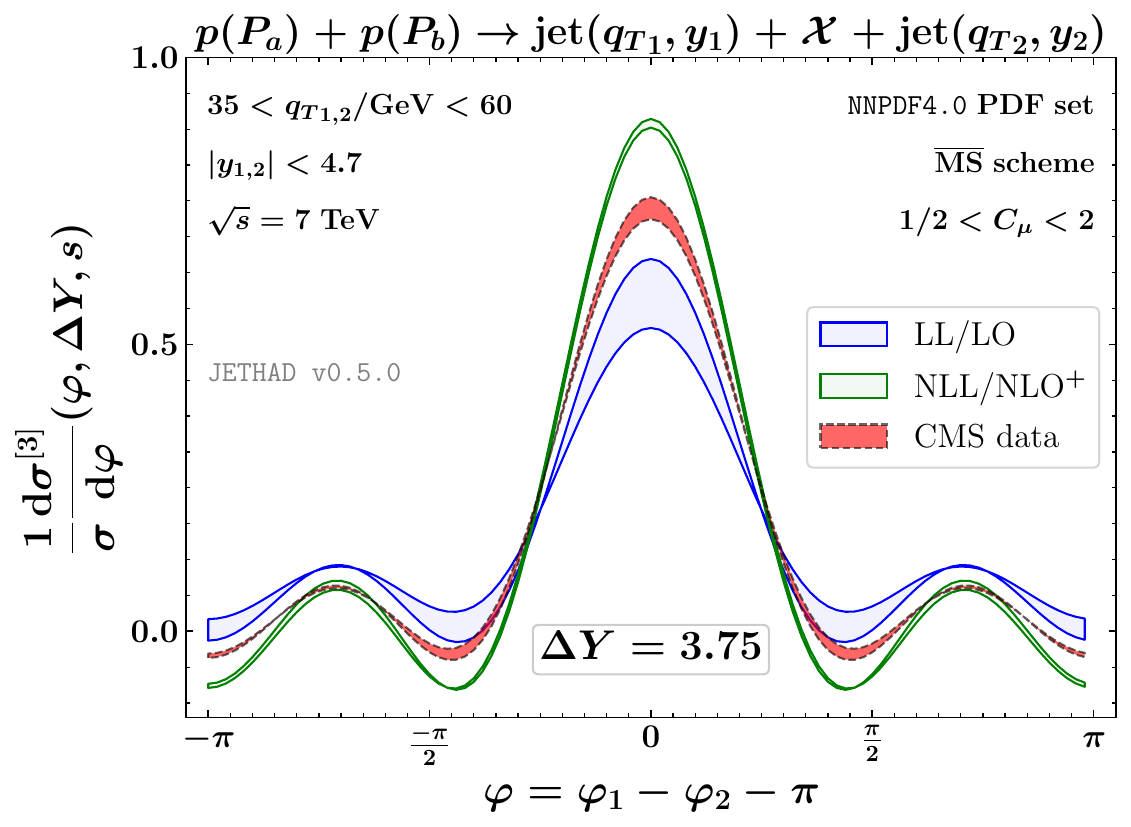}
\includegraphics[scale=0.41,clip]{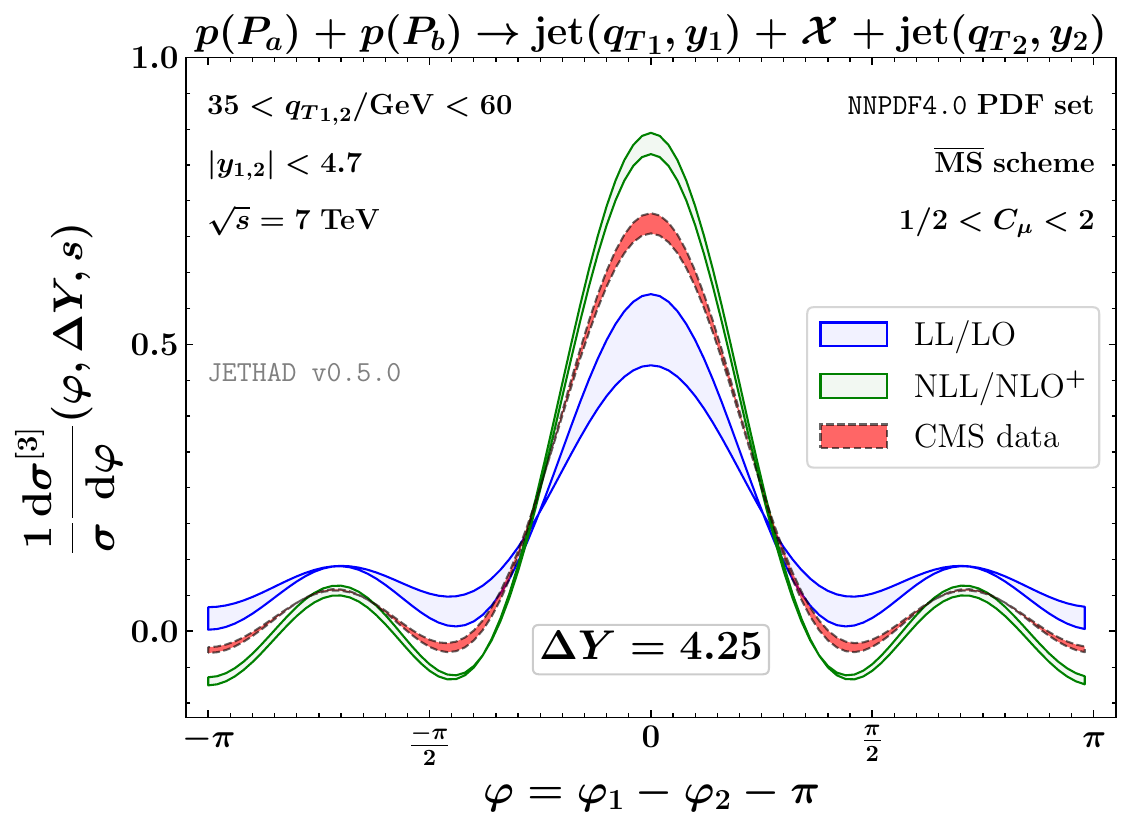}

\includegraphics[scale=0.41,clip]{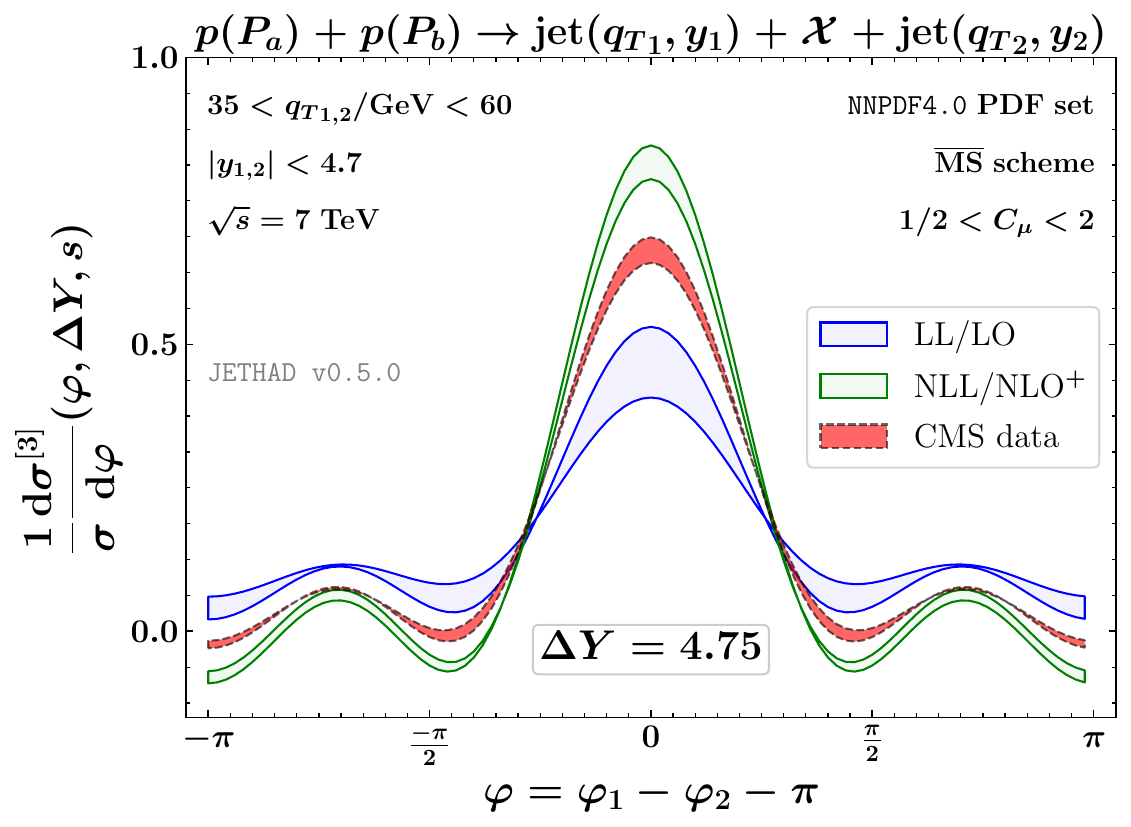}
\includegraphics[scale=0.41,clip]{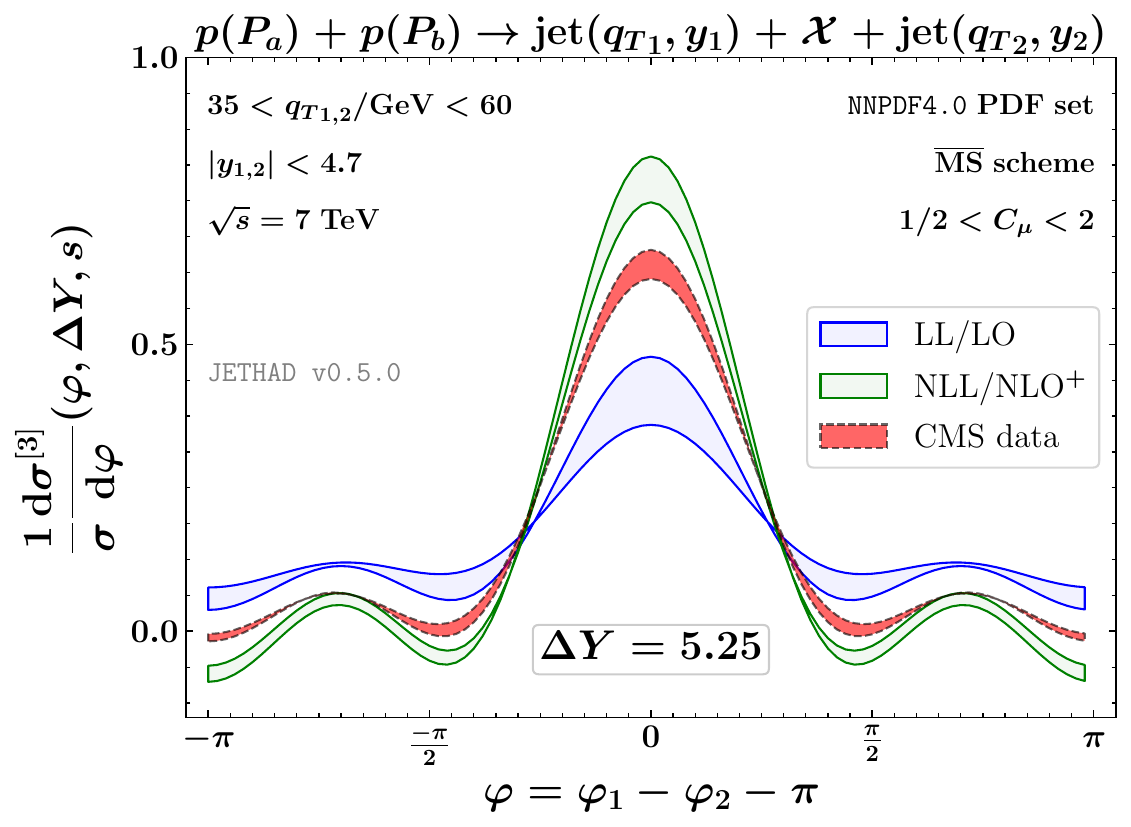}

\caption{Azimuthal distributions truncated at $n^{\max}=3$ for distinct values of $\DY$ and for $\sqrt{s} = 7$ TeV, calculated at natural scales and compared with CMS experimental data (part I). Text boxes inside panels exhibit final-state kinematic cuts. Uncertainty bands of LL/LO and NLL/NLO$^+$ predictions embody the effect of energy-scale variations, while shaded red bands are built in terms of experimental uncertainties.}
\label{fig:phi_tve_1}
\end{figure*}

\begin{figure*}[!t]
\centering

\includegraphics[scale=0.41,clip]{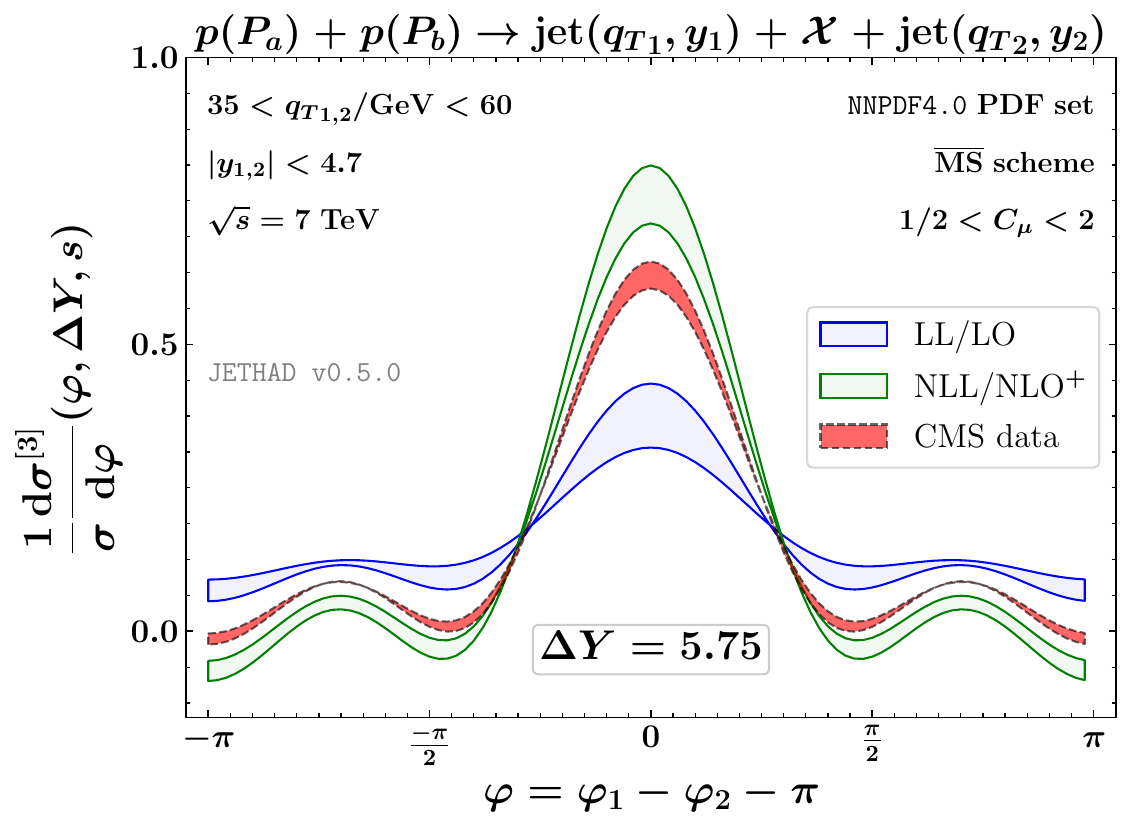}
\includegraphics[scale=0.41,clip]{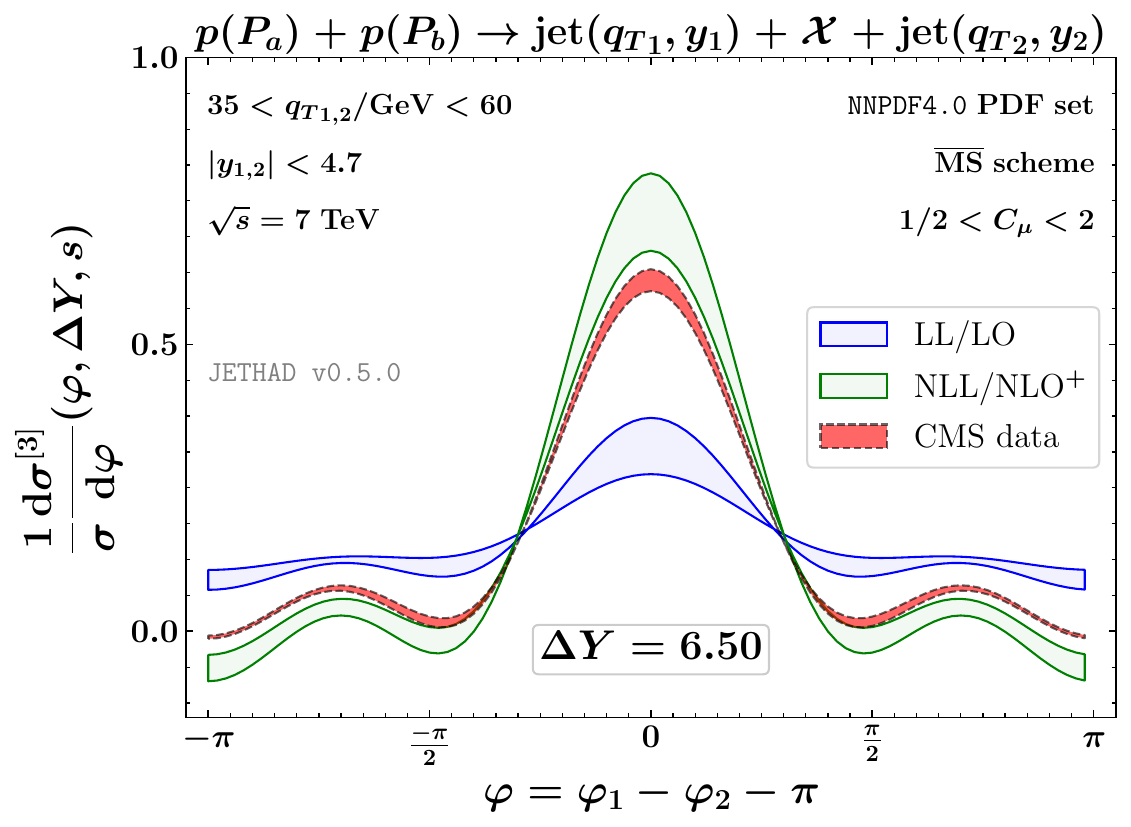}

\includegraphics[scale=0.41,clip]{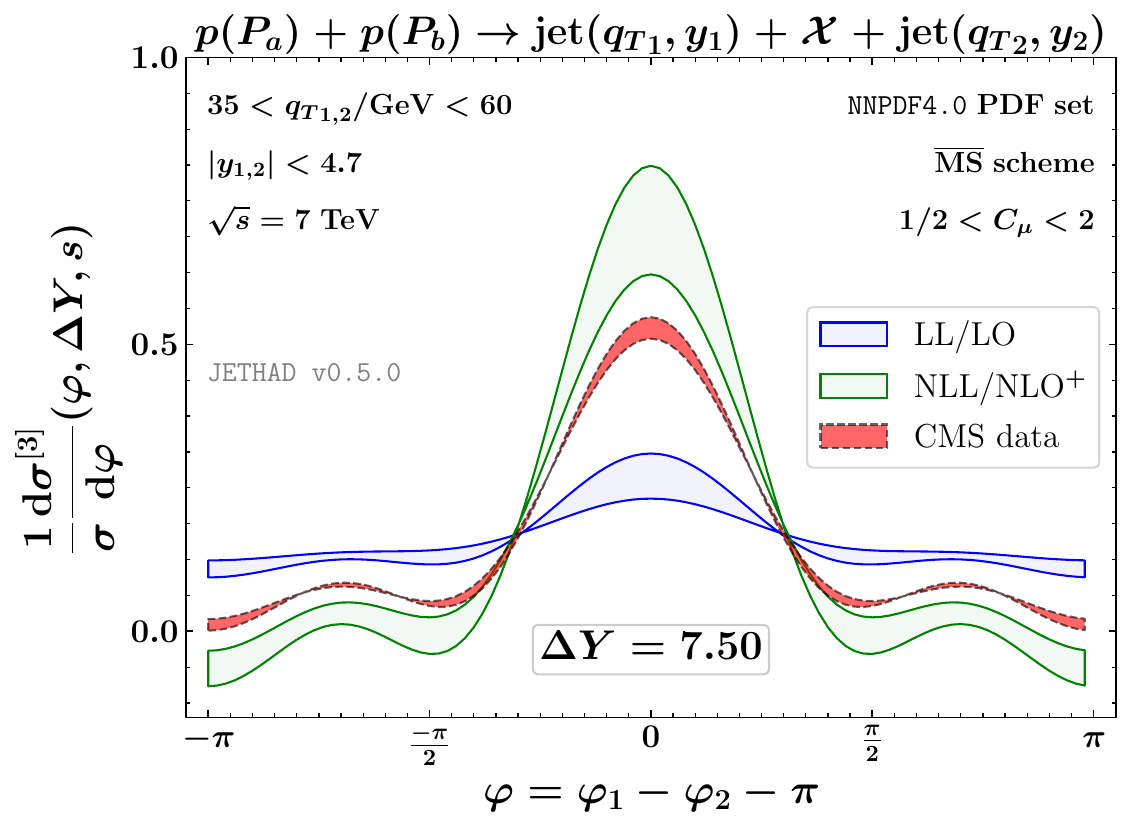}
\includegraphics[scale=0.41,clip]{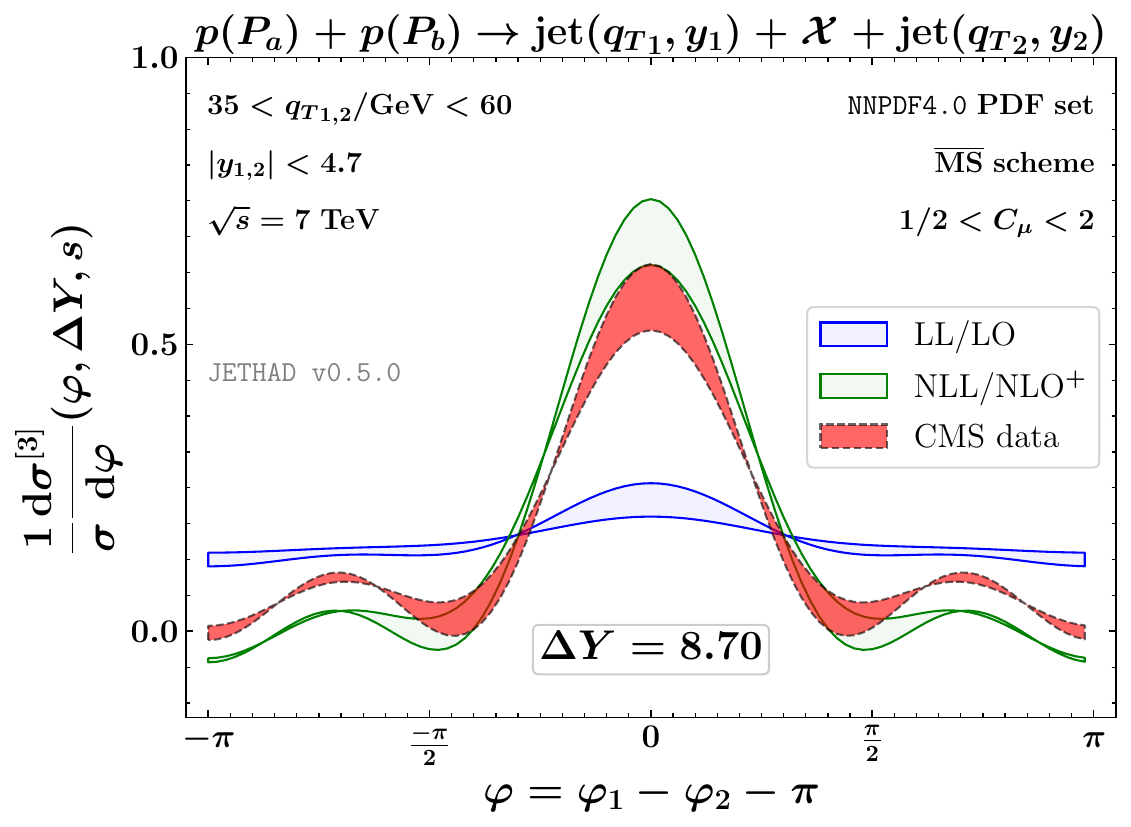}

\caption{Azimuthal distributions truncated at $n^{\max}=3$ for distinct values of $\DY$ and for $\sqrt{s} = 7$ TeV, calculated at natural scales and compared with CMS experimental data (part II). Text boxes inside panels exhibit final-state kinematic cuts. Uncertainty bands of LL/LO and NLL/NLO$^+$ predictions embody the effect of energy-scale variations, while shaded red bands are built in terms of experimental uncertainties.}
\label{fig:phi_tve_2}
\end{figure*}


As predicted by BFKL, the weight of the secondary gluon radiation, emitted with a strong ordering in rapidity, grows with $\DY$. This leads to a downtrend of the jet azimuthal correlation, so that the amount of back-to-back events falls off.
The distance between peaks taken at different $\DY$-values is generally smaller in $\NLLp$ cases. This is in line with the recorrelation pattern generated by NLL contributions both in the kernel and in the jet impact factors (see a related discussion in Section\tref{ssec:Rnm}).
Furthermore, we note that variations of distribution shapes when passing from natural scales to BLM ones (from left to right plots) are more evident at $\LL$ and milder at $\NLLp$.
In particular, these variations are quantitatively smaller then the ones observed in $R_{n0}$ ratios (Fig.\tref{fig:Rn0}).

A clear physical explanation for the fact that azi\-muth\-al-angle distributions suffer from instabilities less than azimuthal coefficients, which are themselves physical observable, is lacking. We argue that some cancellation is at play among different harmonics, producing a compensation of the instabilities affecting the different Fourier terms. The region around $\varphi$=0 is affected by Sudakov resummation effects and the strong peak around zero could make hard the to distinguish between the calculations away from this region. However, this peak is broad enough to possibly leave out a window useful for comparison with data and/or other approaches.

By starting from CMS data available for the first $R_{nm}$ correlations, namely the ones with $n$ ranging from 1 to 3, and $m$ from 0 to 2\tcite{Khachatryan:2016udy},
is it possible to compare the experimental signal, obtained by a suitable combination of those data, with predictions for azimuthal-angle observables derived from distributions in Eq.\eref{dsigma_dphi}, as presented in Section\tref{ssec:hunting_data}.

\subsection{Hunting CMS data at 7 TeV}
\label{ssec:hunting_data}

\begin{figure}[!t]
\centering

\includegraphics[scale=0.43,clip]{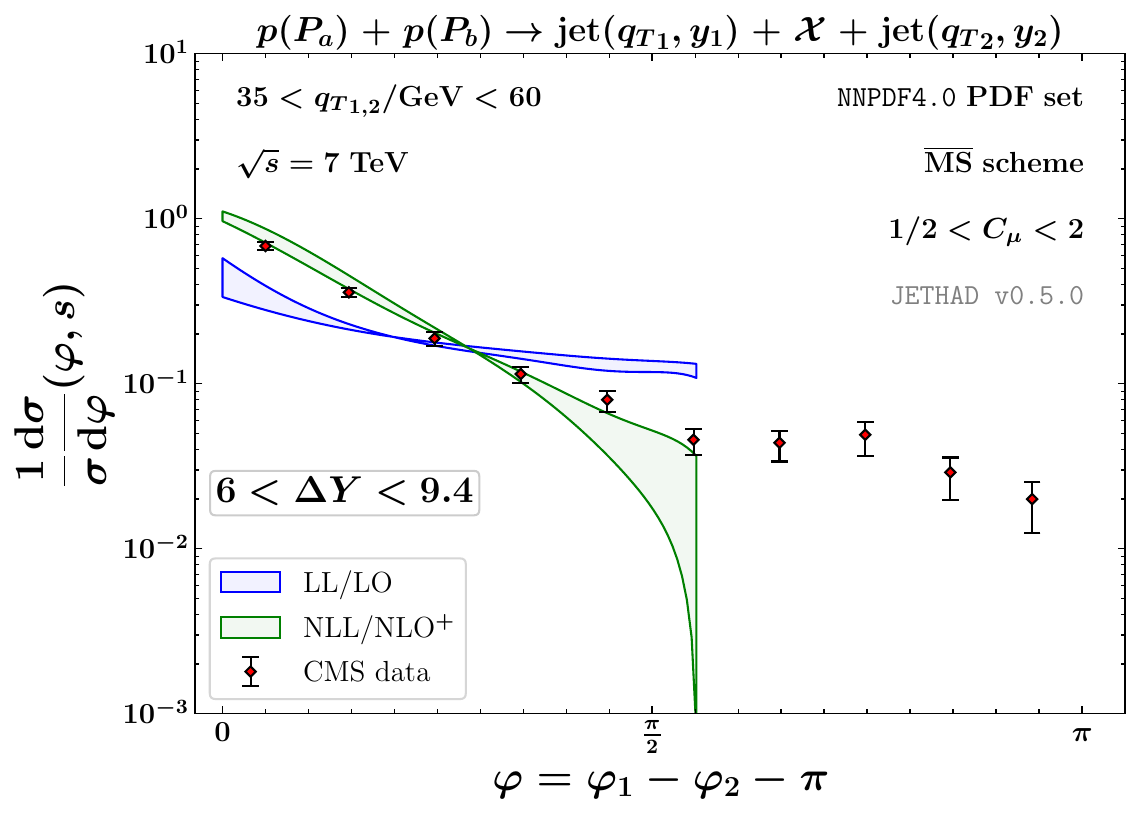}

\caption{Comparison of LL/LO and NLL/NLO$^+$ azimuthal distribution, integrated over the $6 < \DY < 9.4$ range, with CMS data at $\sqrt{s} = 7$ TeV. Text boxes exhibit final-state kinematic cuts. Uncertainty bands embody the effect of energy-scale variations.}
\label{fig:phi-I}
\end{figure}

By summing \emph{\`a la} Fourier the three azimuthal correlations provided by CMS, $R_{10,20,30}$, we can reconstruct the first segment of the experimental signal for the azimuthal distribution. From an operational point of view, we build a \emph{truncated} azimuthal distribution, with $\nu_{\rm [exp]}^{\rm cut} = 3$ dictated by the experimental analysis
\begin{equation}
\label{dsigma_dphi_trunc}
 \frac{1}{\sigma}
 \frac{\drv \sigma^{[3]} (\varphi, \DY, s)}{\drv \varphi} \!=\! \frac{1}{2 \pi}\! \left\{ 1 + 2\! \sum_{n =1 }^{3}\! \cos(n \varphi) R_{n0} \right\} \; .
\end{equation}
In Figs.\tref{fig:phi_tve_1} and\tref{fig:phi_tve_2} we compare $\LL$ and $\NLLp$ predictions for the truncated azimuthal distribution, taken at natural scales, with combined CMS data.
To build the upper (lower) uncertainty bound on CMS data, we considered in Eq.\eref{dsigma_dphi_trunc} the values of the three $R_{n0}$ ratios plus (minus) their uncertainties.
Each panel shows the $\varphi$-shape of our distributions for a given value of $\DY$ taken from the experimental analysis\tcite{Khachatryan:2016udy}.
A visible effect of the truncation in $n$ is the presence of oscillations leading to relative maxima and minima on large-$\varphi$ tails, generated by $1 \le n \le 3$ modes and leading to negative values of the distribution around the corresponding wave troughs.
At small values of $\DY$, the experimental curve (red) stays above the $\LL$ (blue) and below the $\NLLp$ (green) one.
As remarked in Section\tref{ssec:Rnm}, this region crosses the limit of applicability of BFKL. Thus, the LL resummation overestimates the jet azimuthal decorrelation, namely it predicts a lower number of back-to-back events.
At the same time, NLL corrections bring to a very strong recorrelation effect, not compatible with data.
As $\DY$ increases, the experimental curve comes progressively closer to the $\NLLp$ one, up to starting overlap with it when $\DY \ge 6.5$.
Moreover, while these two shapes are always similar, with a clear peak at $\varphi = 0$ which persists also at large values of $\DY$, the peak of the $\LL$ band drops of very fast, with the two jets being almost completely decorrelated when $\DY \ge 7.50$.
As a main outcome, we conclude that the high-energy signal is encoded in LHC data at 7 TeV and can be caught from a NLL BFKL treatment, but not from a pure LL one.
We believe that the agreement between the $\NLLp$ theory and the experiment would be even stronger if data for $R_{n0}$ correlations with $n>3$ were available.
In particular, it would be intriguing to see how the green and the red bands progressively collapse to each other as $\nu_{\rm [exp]}^{\rm cut}$ increases.
Another way to accelerate the convergence between the two curves would be studying $\varphi$-distributions at larger center-of-mass energies, say $\sqrt{s} = 13$ TeV, for which data have been not yet analyzed, unfortunately.
Indeed, the larger is $s$, the faster we move away from endpoints of longitudinal-momentum fractions (see relations in the left part of Eq.\eref{xyp}), where the already mentioned threshold contaminations become relevant.
To better assess this point, we present a comparison with data of LL and NLL azimuthal distributions as in Eq.\eref{dsigma_dphi}, but this time integrated over the rapidity interval, $\DY$.
More in particular, we build $\DY$-integrated $\varphi$-azimuthal distribution
\begin{eqnarray}
\label{dsigma_dphi_int}
 \frac{1}{\sigma}
 \frac{\drv \sigma (\varphi, s)}{\drv \varphi} = \frac{1}{2 \pi} \left\{ 1 + 2 \sum_{n =1 }^{\infty} \cos(n \varphi) R_{n0}^{[\DY\mbox{-}{\rm int}]} \right\}
\end{eqnarray}
as the Fourier sum of azimuthal-correlation ratios
\begin{equation}
 \label{Rn0_int_int}
 R_{n0}^{[\DY\mbox{-}{\rm int}]} =
 \frac{C_n^{[\DY\mbox{-}{\rm int}]}}
 {{C_0^{[\DY\mbox{-}{\rm int}]}}} \; ,
\end{equation}
where $C_n^{[\DY\mbox{-}{\rm int}]}$ are the azimuthal coefficients integrated over the final-state phase space as in Eq.\eref{Cn_int} and also over $\DY$.
A limited amount of data for $\DY$-integrated $\varphi$-distri\-butions was collected by CMS at $\sqrt{s} = 7$~TeV and for symmetric transverse-momentum windows. As shown in Fig.~1 of Ref.\tcite{Khachatryan:2016udy}, these data fall into three $\DY$-bins. For the sake of clarity, we consider just the $6 < \DY < 9.4$ one, namely where distinctive high-energy signatures are expected due to large rapidity intervals. We compare resummed predictions with CMS data from Ref.\tcite{Khachatryan:2016udy} divided by a factor two, due to the fact that in our analysis only $\DY > 0$ values are considered, namely when the first jet is always more forward than the second one.
From the inspection of results in Fig.\tref{fig:phi-I} (see also Refs.\tcite{Ducloue:2013bva,Ducloue:2014bpa} for a quite similar study), it emerges that a pure LL treatment does not catch data.
Conversely, NLL predictions are in a fair agreement with data in the $\varphi \lesssim \pi/2$ region, namely where the jets are emitted (almost) back to back, or their transverse-momentum imbalance is small. Then, the NLL description significantly worsen as $\varphi$ grows, up to reaching unphysical values not shown in our plot. Here we enter a region where the two observed transverse momenta are different, which is not an optimal condition for BFKL. Therefore, other resummation effects, not caught by our hybrid factorization, could be relevant.

\section{Future perspectives}
\label{sec:conclusions}

We compared predictions for Mueller-Navelet jet rapidity and azimuthal-angle differential distributions with data collected by CMS at $\sqrt{s} = 7$ TeV.
We made use of the hybrid collinear and high-energy factorization as specifically designed for inclusive Mueller-Navelet emissions\tcite{Colferai:2010wu,Ducloue:2013hia,Ducloue:2013bva,Caporale:2013uva,Caporale:2014gpa,Celiberto:2015yba,Caporale:2015uva,Celiberto:2016ygs,Celiberto:2016vva}, where the collinear description in terms of PDFs is supplemented by the BFKL resummation of NLL energy logarithms.
To be close to standard methodologies widely employed for precision studies of high-energy reactions done by the hands of fixed-order calculations as well as of other resummations, we gauged the size of higher-order corrections by generating scale-variation driven uncertainty bands. This procedure was introduced in the context of semihard process through a study on Higgs-plus-jet correlations done by our Group\tcite{Celiberto:2020tmb}, but it turns out to be novel for Mueller-Navelet jets, which represents the ``mother'' of the forward-plus-backward subclass of semihard reactions.

We provided an evidence that considering a\-zi\-mu\-thal-angle dependent distributions, calculated as a Fou\-rier sum of azimuthal correlations and carrying the complete high-energy signal emerging from all con\-for\-mal-spin modes, allows us to dampen instabilities rising when inclusive tags of light objects are theoretically investigated at the natural scales provided by kinematics.
We came out with a clear indication that the NLL BFKL treatment of $\varphi$-distributions becomes more and more valid as $\DY$ grows, and it permits us to catch the core high-energy dynamics emerging from CMS data collected so far. In particular, we observed that azimuthal-angle distributions at the NLO are more stable and less dependent on the choice of the renormalization and factorization scales, with respect to the single azimuthal coefficients from which they are built.

In view of these considerations, we warmly suggest experimental collaborations to include in forthcoming analyses at $\sqrt{s} = 13$ TeV $(i)$ a dedicated study of Mueller-Navelet azimuthal distributions as well as $(ii)$ an asymmetric-window selection for the observed transverse momenta.
We believe that combining these ingredients is relevant to accelerate our progresses in unveiling the presence of high-energy dynamics in Mueller-Navelet final states, and to better disengage the BFKL signal from the DGLAP background.

Concerning future developments on the theory side, we are aware that our path toward reaching the precision level in the description of high-energy Mueller-Navelet emissions and, more in general, of semihard reactions, moves through a robust enhancement of our hybrid factorization.
Starting from our hybrid factorization, we plan the development of a \emph{multilateral} and unified formalism where distinct resummation mechanisms, in particular BFKL, threshold and Sudakov \cite{Mueller:2012uf,Mueller:2013wwa,Balitsky:2015qba,Marzani:2015oyb,Mueller:2015ael,Xiao:2018esv} ones, are simultaneously embodied.

\section*{Acknowledgments}
\label{sec:acknowledgments}

We thank Dmitry Yu. Ivanov and Marco Bonvini for insightful discussions on progresses and challenges in the implementation of the BFKL-plus-threshold double logarithmic resummation.
F.G.C. acknowledges support from the INFN/NIN\-PHA project and thanks the Universit\`a degli Studi di Pavia for the warm hospitality.
A.P. acknowledges support from the INFN/QFT@\-COL\-LI\-DERS project.

\bibliographystyle{apsrev}
\bibliography{references}

\end{document}